\documentclass[pdflatex,sn-nature]{sn-jnl}

\usepackage{graphicx}%
\usepackage{multirow}%
\usepackage{amsmath,amssymb,amsfonts}%
\usepackage{amsthm}%
\usepackage{mathrsfs}%
\usepackage[title]{appendix}%
\usepackage{xcolor}%
\usepackage{textcomp}%
\usepackage{manyfoot}%
\usepackage{booktabs}%
\usepackage{algorithm}%
\usepackage{algorithmicx}%
\usepackage{algpseudocode}%
\usepackage{listings}%
\usepackage{bold-extra}

\usepackage{bold-extra}
\theoremstyle{thmstyletwo}%

\theoremstyle{thmstylethree}%

\newcommand{\ion}[2]{{#1}\,{\scshape{#2}}}

\begin{document}


\title[Discovery of a new transitional type of evolved massive stars]{Discovery of a new transitional type of evolved massive stars with hard ionizing flux}


\author*[1]{\fnm{Andreas A.C.} \sur{Sander}}\email{andreas.sander@uni-heidelberg.de}
\author[1]{\fnm{Roel R.} \sur{Lefever}}\email{roel.lefever@uni-heidelberg.de}
\author[1]{\fnm{Joris} \sur{Josiek}}\email{joris.josiek@uni-heidelberg.de}
\author[2]{\fnm{Erin R.} \sur{Higgins}}\email{Erin.Higgins@armagh.ac.uk}
\author[3,9]{\fnm{Raphael} \sur{Hirschi}}\email{r.hirschi@keele.ac.uk}
\author[4]{\fnm{Lidia M.} \sur{Oskinova}}\email{lida@astro.physik.uni-potsdam.de}
\author[4]{\fnm{Daniel} \sur{Pauli}}\email{danielpauli1997@gmail.com}
\author[4]{\fnm{Max} \sur{Pritzkuleit}}\email{maxpri@astro.physik.uni-potsdam.de}
\author[5]{\fnm{John S.} \sur{Gallagher}}\email{ajayg3@gmail.com}
\author[4]{\fnm{Wolf-Rainer} \sur{Hamann}}\email{wrh@astro.physik.uni-potsdam.de}
\author[6,7]{\fnm{Ilya} \sur{Mandel}}\email{ilya.mandel@monash.edu}
\author[1]{\fnm{Varsha} \sur{Ramachandran}}\email{vramachandran@uni-heidelberg.de}
\author[8]{\fnm{Tomer} \sur{Shenar}}\email{tshenar@tauex.tau.ac.il}
\author[4]{\fnm{Helge} \sur{Todt}}\email{htodt@astro.physik.uni-potsdam.de}
\author[2]{\fnm{Jorick S.} \sur{Vink}}\email{Jorick.Vink@armagh.ac.uk}

\affil*[1]{\orgname{Zentrum f{\"u}r Astronomie der Universit{\"a}t Heidelberg}, \orgdiv{Astronomisches Rechen-Institut}, \orgaddress{\street{M{\"o}nchhofstr. 12-14}, \postcode{69120} \city{Heidelberg}, \country{Germany}}}

\affil[2]{\orgname{Armagh Observatory and Planetarium}, \orgaddress{\street{College Hill}, \city{Armagh} \postcode{BT61 9DG}, \state{Northern Ireland}, \country{UK}}}

\affil[3]{\orgdiv{Astrophysics Group}, \orgname{Keele University}, \orgaddress{\street{Keele}, \city{Staffordshire} \postcode{ST5 5BG}, \country{UK}}}

\affil[4]{\orgdiv{Institut f{\"u}r Physik \& Astronomie}, \orgname{Universit{\"a}t Potsdam}, \orgaddress{\street{Karl-Liebknecht-Str.\ 24/25}, \postcode{14476} \city{Potsdam}, \country{Germany}}}

\affil[5]{\orgdiv{Department of Astronomy}, \orgname{University of Wisconsin}, \orgaddress{\city{Madison}, \state{WI}, \country{USA}}}

\affil[6]{\orgdiv{School of Physics and Astronomy}, \orgname{Monash University}, \orgaddress{\city{Clayton} \state{VIC} \postcode{3800}, \country{Australia}}}

\affil[7]{\orgname{ARC Center of Excellence for Gravitational-wave Discovery (OzGrav)}, \orgaddress{\city{Melbourne}, \country{Australia}}}

\affil[8]{\orgdiv{The School of Physics and Astronomy}, \orgname{Tel Aviv University}, \orgaddress{\city{Tel Aviv} \postcode{6997801}, \country{Israel}}}

\affil[9]{\orgdiv{Kavli IPMU (WPI)}, \orgname{University of Tokyo},  \orgaddress{\street{5-1-5 Kashiwanoha}, \city{Kashiwa} \postcode{277-8583}, \country{Japan}}}

\abstract{%
  Wolf-Rayet (WR)
  stars are the evolved descendants of the most massive stars and show emission-line dominated spectra formed in their powerful stellar winds. Marking the final evolution stage before core collapse, the standard picture of WR stars has been that they evolve through three well-defined spectral subtypes known as WN, WC, and WO. Here, we present a detailed analysis of five objects that defy this scheme, demonstrating that WR stars can also evolve directly from the WN to the WO stage. Our study reveals that this direct transition is connected to low metallicity and weaker winds. The WN/WO stars and their immediate WN precursors are hot and emit a high flux of photons capable of fully ionizing helium. The existence of these stages unveil that high mass stars which manage to shed off their outer hydrogen layers in a low-metallicity environment can spend a considerable fraction of their lifetime in a stage that is difficult to detect in integrated stellar populations, but at the same time yields hard ionizing flux.
  The identification of the WN to WO evolution path for massive stars has significant implications for understanding the chemical enrichment and ionizing feedback in star-forming galaxies, in particular at earlier cosmic times.
}%

\maketitle

\section{Introduction}\label{sec:intro}

Classical Wolf-Rayet (WR) stars are helium-burning massive stars whose outer hydrogen layers have been partially or completely stripped off \cite{Conti1975,Crowther2007}. Owing to the high ratio between their stellar luminosity $L$ and their mass $M$, WR stars are able to drive very strong stellar winds that lead to emission-line spectra, giving them their characteristic spectral appearance \cite{VinkdeKoter2002,GraefenerHamann2008,Sander+2020}. WR stars are further classified into the WN, WC, and WO subtypes characterized by the presence of prominent nitrogen (N), carbon (C), and oxygen (O) lines \cite{Smith+1996,Crowther+1998}. With their stripped-envelope nature, WR stars provide a rare opportunity to directly observe layers with abundances that are otherwise hidden in the interiors of the massive stars, thereby marking important test cases for the fundamental theory of stellar evolution.
The clear visibility of nuclear processed material at the surface of WR stars early on established the idea that WN, WC, and WO stars also form an evolutionary sequence \cite{Gamow1943} with WO stars marking
the final evolution stage before core collapse. The derived surface abundances and post-processing of evolution tracks provided further support to this scenario \cite{Langer1991,Meynet+1994,Groh+2014,Aadland+2022}. The spectral transition between the stages is usually seen as smooth due the discovery of objects with hybrid-character between the WN and WC stage, denoted as WN/WC \cite{MasseyGrove1989,Crowther+1995,Shara+2016,Zhang+2020,Hillier+2021}, as well as comparison studies between WC and WO-type stars \cite{Aadland+2022wc,Aadland+2022}.

In this work, we present two objects that clearly directly defy the standard picture of a WN to WC to WO transition: M33WR\,206 and M31WR\,99-1 show spectra that combine WN and WO-type features. Previously, these objects were classified as WN2 \cite{NeugentMassey2011} and WO \cite{NeugentMassey2023}, respectively. For the first time, we perform a detailed spectral analysis, revealing that not only their spectral appearance but also their stellar properties align with a direct transition from the WN to the WO stage. The quantitative comparison with other objects of a specific WN spectral subtype -- the WN2 class -- further indicates that all known members of the WN2 subclass are precursors of the newly discovered WN/WO-transition stage. Characterized by hot temperatures and -- compared to other WR subtypes -- weaker winds, the WN/WO and WN2 stars are emitting a large number of photons with wavelengths below the \ion{He}{ii} ionizing edge where the winds of most other WR stars are opaque. Our atmosphere analysis combined with evolutionary calculations reveals that this stage is reached by massive stars at low metallicity which manage to completely lose their hydrogen. The direct transition from the WN to the WO stage further explains the scarcity of WC stars in low metallicity populations \cite{Crowther2007,Monreal-Ibero+2017} and the absence of the corresponding imprint in integrated galaxy spectra, while at the same time providing strong sources of \ion{He}{ii} ionizing flux \cite{Kehrig+2018,Mayya+2023}. The existence of a WR stage with hard ionizing flux at low metallicity could also mark a major puzzle piece in explaining highly ionized nebular emission lines seen in various high-redshift galaxies detected by JWST \cite{Carniani+2024,Castellano+2024,DEugenio+2024}.

\section{Stellar parameters and abundances}

Using newly obtained HST COS spectra as well as archival UV and optical \cite{Hamann+1993,NeugentMassey2011,NeugentMassey2023} data complemented with photometry (see methods part for details), we study the full sample of WN2 and newly discovered WN/WO stars with sufficient spectral coverage to perform a multi-wavelength quantitative spectral analyses. Our sample comprises the WN/WO stars M33WR\,206 and M31WR\,99-1 as well as the WN2 stars WR\,2, BAT99\,2, and BAT99\,5, the only sources of this kind with available UV spectra. A UV and optical comparison of their spectra is given in Figs.\,1 and 2, respectively. 
In total, our sample covers four different galaxies, namely the Milky Way, the Large Magellanic Cloud (LMC), the Andromeda Galaxy (M31) and the Triangulum Galaxy (M33).
For our quantitative analysis, we employ the stellar atmosphere code PoWR \cite{Graefener+2002,HamannGraefener2003,Sander+2015}, which solves the statistical equations and the radiative transfer in an expanding environment outside of local thermodynamic equilibrium (non-LTE). In this work, we further use the PoWR$^\textsc{hd}$ branch to run hydrodynamically-consistent models \cite{Sander+2020,Sander+2023} in which the wind velocity stratification and the mass-loss rate are consistently predicted from the stellar parameters. Getting the first self-consistent spectral fits for early-type WN and WN/WO stars marks a pioneering effort in itself, ensuring that our derived temperatures and ionizing fluxes are not subject to the considerable parameter degeneracies arising from the optically thick winds of WR stars \cite{HamannGraefener2004,Lefever+2023}. 

The combined features of a WN and WO-type star could suggest that our sources stem from a superposition of a WN and WO star, either due to binarity or crowding. As we discuss in more detail in the methods part, this is highly unlikely as the combined features of the two types of stars do not yield the spectral appearance we observe for M33WR\,206 and M31WR\,99-1 when considering known representatives of each class.

\subsection{Stellar parameters}

\begin{figure}
    \centering
\includegraphics[width=0.85\columnwidth]{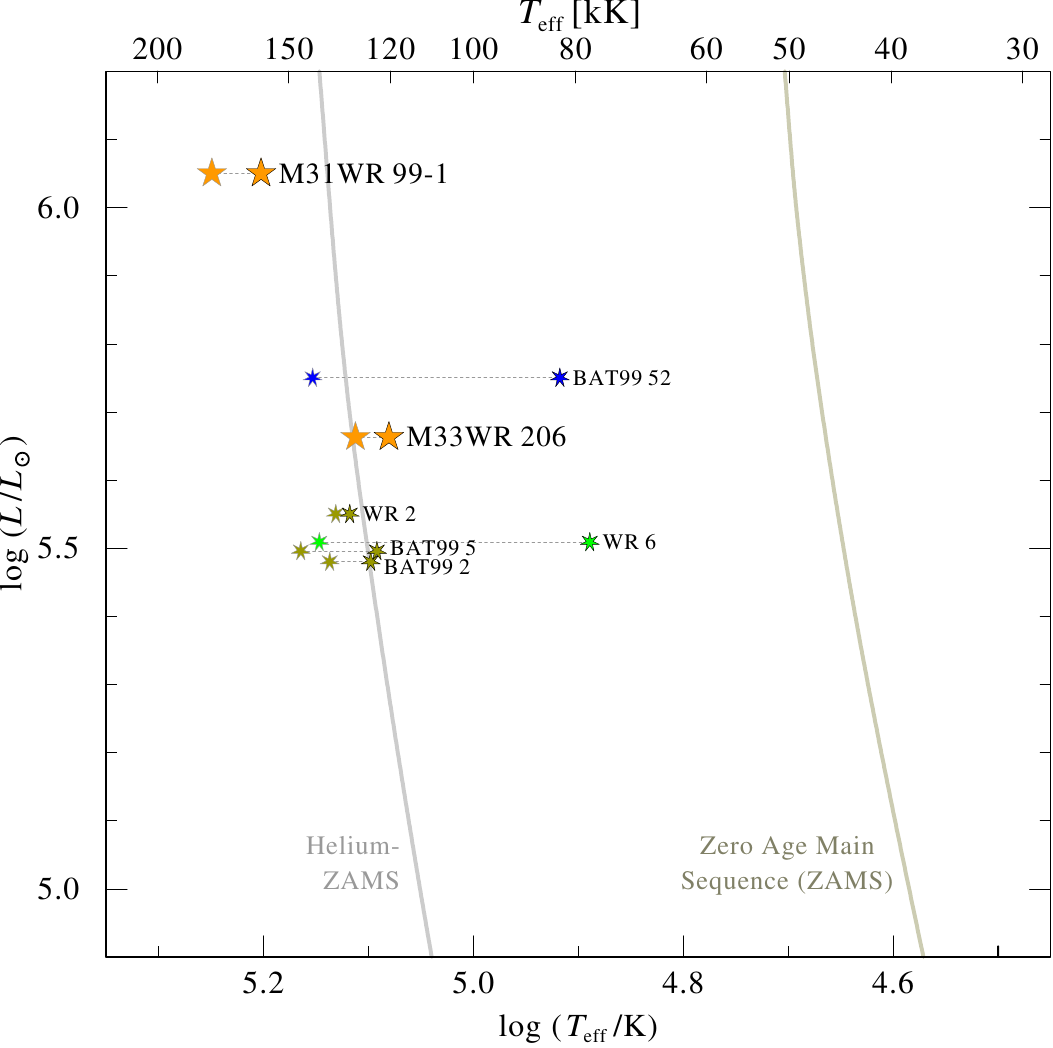}
  \caption{Hertzsprung Russell Diagram (HRD) with the positions of the newly discovered WN/WO stars (orange) compared to three other WN2 stars ({WR\,2}, {BAT99\,2}, {BAT99\,5}, olive) as well as typical examples of a WN (green) and WC star (blue) with stronger winds. For each star, the standard effective temperature at $\tau = 2/3$ (solid border symbols) is shown together with the effective temperature corresponding to the onset of the stellar wind ($R_\mathrm{crit}$, light border symbols). Both temperatures are connected by a dashed line to guide the eye. \label{fig:3}}
\end{figure}

All of our sample stars have luminosities typical for massive WR stars but show very high effective temperatures between $120$ and $160\,$kK. Their atmospheres are slightly extended with their photospheric radii being located in the supersonic wind. This is illustrated in Fig.\,\ref{fig:3} where we plot the position of the stars in the Hertzsprung-Russell diagram (HRD), showing for each star both the effective temperature $T_\mathrm{eff}(R_\mathrm{crit})$ at the onset of the wind $R_\mathrm{crit}$ as well as the presumed effective temperature $T_\mathrm{eff}(R_{2/3})$ inferred from the radius $R_{2/3}$ where the radiation leaves the star. For comparison, we also plot the two temperatures for a typical WN (WR\,6) and WC (BAT99\,52) star in Fig.\,\ref{fig:3}, which were also analysed with hydrodynamically-consistent models. As evident, the extension of the optically thick wind regime for the WN/WO and WN2 stars is considerably smaller than for the later-type WN or WC stars. This more compact appearance is typical for the earliest WN stars and WO stars and aligns with generally lower mass-loss rates compared to other WR stars \cite{GraefenerVink2013,Sander+2020}. Indeed, we derive mass-loss rates of $10^{-5.5}$ and $10^{-4.95}\,M_\odot\,\mathrm{yr}^{-1}$ for M33WR\,206 and M31WR\,99-1, respectively, which are high, but far from what can be reached by some of the later-type WR stars \cite{Grassitelli+2018,Sander+2023}.

\subsection{Nitrogen and Oxygen abundances}

Both M33WR\,206 and M31WR\,99-1 show nitrogen lines in their spectra, which are not observed for any known WO star and are also very rare in WC stars \cite{Hillier+2021}. The derived nitrogen abundances are typical for WN stars in their corresponding host galaxies. Thus, the spectra of M33WR\,206 and M31WR\,99-1 are typical for WN stars of the earliest subtype, except for a strong oxygen emission line characteristic for WO stars. Compared to known WO-type stars which have between 7\% and 25\% of oxygen (by mass) in their atmosphere \cite{Tramper+2015,Aadland+2022}, the oxygen abundance required to reproduce the WN/WO spectral appearance is very small with only 0.5\% by mass for M33WR\,206 and 1.5\% for M31WR\,99-1. In contrast, WN2 stars such as WR\,2, BAT99\,2 and BAT99\,5 have similar nitrogen abundances but much lower oxygen abundances with the highest upper limit being $0.01\%$ in the case of WR\,2. Therefore, we can conclude that M33WR\,206 and M31WR\,99-1 are genuinely in a transition stage between the WN and the WO spectral appearance, which we label as WN/WO, analogous to other spectral transition notations. A full illustration of the spectral fits for all five targets is given in the methods part.

\subsection{Carbon and Neon abundances}

\begin{figure}
    \centering
\includegraphics[width=\columnwidth]{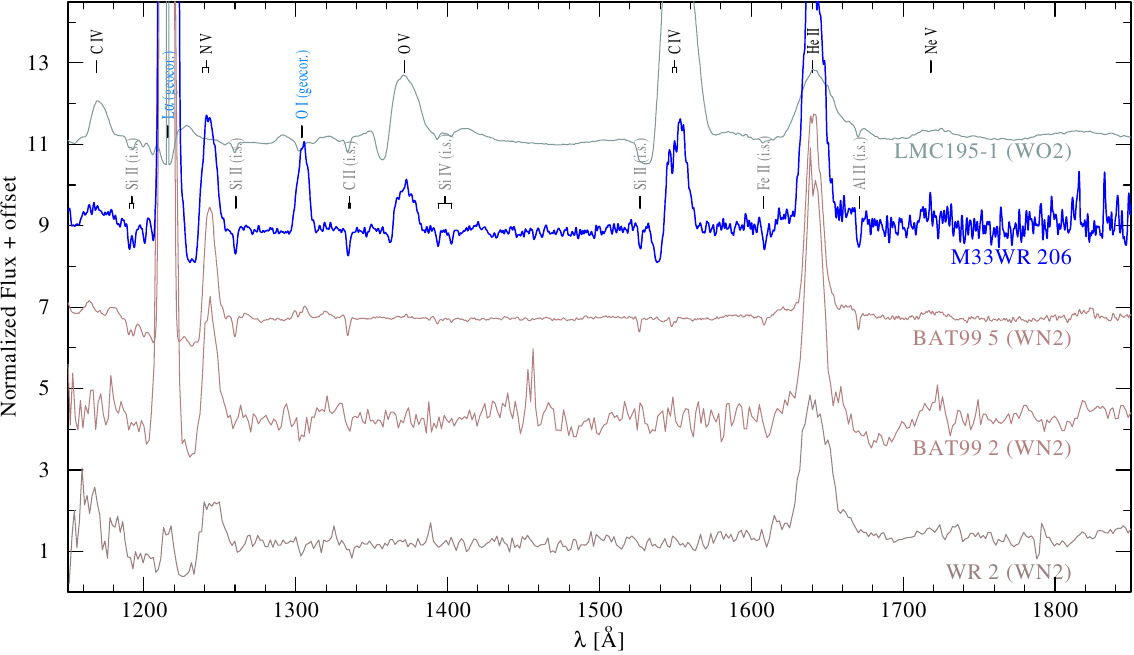}
\caption{Normalized UV spectrum of the WN2/WO1 star {M33WR\,206} compared to three other WN2 stars in the Milky Way ({WR\,2}) and the LMC ({BAT99\,2}, {BAT99\,5}) as well as the WO2 star {LMC195-1}. The normalized spectra have been shifted along the ordinate for clearer distinction between the different objects. \label{fig:1}}
\end{figure}

Both newly identified WN/WO stars display strong oxygen lines but only weak carbon lines in their optical spectra. Nonetheless, our spectroscopic analysis reveals that their carbon abundances ($2\%$ and $5\%$ by mass for M33WR\,206 and M31WR\,99-1, respectively) in the WN/WO stars are notably higher than their oxygen abundances.
This demonstrates that although most known WO stars are actually further evolved than WC stars \cite{Aadland+2022}, the WO spectral appearance as such is largely a temperature effect.
Consequently, the newly identified WN/WO stars can be interpreted as hotter counterparts of the, also rare, WN/WC-transition type \cite{ContiMassey1989,Shara+2016,Zhang+2020} which we further illustrate in Fig.\,\ref{fig:wrtempseq}. In WN/WO atmospheres, carbon is highly ionized and thus barely visible in the optical spectrum. For M33WR\,206, where UV spectroscopy is available, carbon shows a prominent feature in the far UV (see Fig.\,\ref{fig:1}), as do both oxygen and nitrogen. For BAT99\,5, the good quality UV spectrum shows a weak \ion{C}{iv} wind profile, which we utilize to constrain an absolute value for the carbon abundance.

Another elemental abundance -- mainly accessible in the UV for our objects -- is neon. In Fig.\,\ref{fig:1}, one can see a comparably weak but noticeable UV emission feature around $1720\,$\AA, which is usually attributed to \ion{N}{iv} $\lambda\lambda 1719$-$1722$ and successfully reproduced in stellar atmosphere models for later-type WN stars \cite{Herald+2001,Hamann+2006,Hainich+2015}. However, the ionization in the atmospheres of WN2 stars is so high that \ion{N}{iv} is not significantly populated and all \ion{N}{iv} features in the optical and UV are absent. The explicit inclusion of neon into the model atmospheres reveals that the UV emission feature around $1720\,$\AA\ arises from \ion{Ne}{v} $\lambda 1718$. A neon mass fraction of about $0.5\%$ is necessary to reproduce the observed line strength in M33WR\,206.
While M31WR 99-1 has no available UV spectra, neon is also detected in the UV spectra of the WN2 stars -- WR\,2, BAT99\,2, and BAT99\,5. 

The two WN2 stars with IUE spectral coverage (WR\,2 and BAT99\,2) further show additional neon imprints at longer UV wavelengths ($\sim$$2230\,$\AA). While the precise abundance values are subject to some uncertainty due to the limited signal-to-noise of the underlying data in this wavelength region, the distinct visibility of the neon features in the models often requires abundances slightly higher than usually expected for WN stars. If we take these number at face value, they could indicate that WN2 stars are more evolved than other WN stars and close to the discovered WN/WO transition. More evidence for a general connection between WN2 and WN/WO stars is yet given by the spectrum of the WN2 star BAT99\,5, which shows a small but noticeable excess around 3811\,\AA\ which we can reproduce as \ion{O}{VI} emission with our atmosphere model. Given that optical \ion{O}{VI} emission is never seen in any later-type WN stars, but in early WC and WO stars, the case of BAT99\,5 clearly links the WN2 stars to the WN/WO stars. Yet, the derived oxygen abundance is not in line with a possibly enhanced neon abundance. The full set of abundances for all the targets and their uncertainties are provided in Table 1.

\section{Evolutionary status of the WN/WO-stars and a new perspective on the WN2 class}

\begin{table*}
   \caption{Derived stellar parameters for the analyzed early-type WR stars \label{tab:results}. 
   The surface abundances are given as mass fractions $X_i$ for each element $i$. $v_\mathrm{cl}$ denotes the characteristic velocity of the clumping onset \cite{HillierMiller1999} and $D_\infty$ denotes the maximum clumping factor at infinity. The mechanical luminosity is calculated as $L_\mathrm{mech} = 0.5 \dot{M} v_\infty^2$.}
   \centering
    \renewcommand{\arraystretch}{1.3}
    \begin{tabular}{l c c c c c}
      \hline 
         Parameter                   &  M33WR\,206  & WR\,2  & BAT99\,2  &  BAT99\,5  &  M31WR\,99-1 \\ 
                                     &    WN2/WO1   &   WN2  &    WN2    &      WN2   &   WN2/WO1    \\ 
      \hline
       $T_\ast$ ($\tau=100$) [kK]  &   $135^{+5}_{-3}$       & $140^{+6}_{-6}$        &    $142^{+6}_{-6}$      &  $150^{+4}_{-6}$  & $180^{+15}_{-30}$ \\
       $T_\mathrm{eff}(R_\mathrm{crit})$ [kK]  &   $129^{+5}_{-3}$       & $135^{+6}_{-6}$        &    $137^{+6}_{-6}$      &  $146^{+4}_{-6}$  & $177^{+15}_{-30}$ \\
       $T_{2/3}$ ($\tau=2/3$) [kK]              &  $121^{+5}_{-8}$        & $131^{+6}_{-6}$        &    $125^{+4}_{-5}$      &  $126^{+9}_{-8}$ & $159^{+6}_{-22}$ \\
       $R_\ast$ [$R_\odot$]        &  $1.22^{+0.06}_{-0.04}$ & $1.02^{+0.01}_{-0.02}$ & $0.91^{+0.08}_{-0.03}$ & $0.84^{+0.11}_{-0.03}$ & $1.09^{+0.29}_{-0.06}$ \\
       $R_\mathrm{crit}$ [$R_\odot$] &  $1.34^{+0.09}_{-0.08}$ & $1.09^{+0.04}_{-0.04}$ & $1.03^{+0.09}_{-0.07}$  & $0.88^{+0.15}_{-0.04}$ & $1.12^{+0.33}_{-0.09}$\\
       $R_{2/3}$ [$R_\odot$]       &   $1.53^{+0.21}_{-0.08}$ & $1.16^{+0.04}_{-0.04}$ & $1.17^{+0.17}_{-0.08}$  &  $1.19^{+0.06}_{-0.17}$ &  $1.39^{+0.36}_{-0.06}$ \\
       $\log L/L_\odot$            &   $5.65^{+0.05}_{-0.05}$ & $5.55^{+0.1}_{-0.05}$  &   $5.48^{+0.10}_{-0.10}$     &  $5.5^{+0.05}_{-0.05}$  &  $6.05^{+0.10}_{-0.15}$   \\
       $M/M_\odot$                 &   $17.9^{+1.7}_{-1.9}$  & $17.4^{+2.8}_{-3.2}$    &   $13.7^{+3.3}_{-2.6}$    &  $13.2^{+2.4}_{-2.9}$  & $33.7^{+17.0}_{-7.0}$  \\
       $\log \left(\dot{M}\,[M_\odot\,\mathrm{yr}^{-1}]\right)$  &  $-5.5^{+0.10}_{-0.05}$ & $-5.65^{+0.13}_{-0.10}$ & $-5.6^{+0.15}_{-0.15}$  & $-5.43^{+0.10}_{-0.10}$      & $-4.95^{+0.22}_{-0.22}$ \\
       $\log \left(\dot{M}_\mathrm{t}\,[M_\odot\,\mathrm{yr}^{-1}]\right)$  &  $-4.70^{+0.05}_{-0.11}$  &   $-5.00^{+0.10}_{-0.10}$   &  $-4.80^{+0.10}_{-0.10}$ &  $-4.81^{+0.10}_{-0.10}$ & $-5.23^{+0.05}_{-0.06}$ \\
       $v_\infty$ [km\,s$^{-1}$] &   $2400^{+300}_{-300}$ & $3400^{+400}_{-200}$  &    $1800^{+300}_{-200}$  & $1700^{+300}_{-100}$ &  $5500^{+200}_{-500}$  \\
       $D_\infty$                     &   $50^{+50}_{-30}$  & $50^{+50}_{-30}$  &   $20^{+10}_{-5}$  &  $10^{+5}_{-6}$  &  $10^{+20}_{-6}$ \\
       $v_\mathrm{cl}$ [km\,s$^{-1}$] &  180  & 250 &  250  &  500 &  150  \\
       $X_\mathrm{He}$ (mass fr.)     &  $0.965$                   &  $0.979$                  &  $0.984$                  &  $0.987$  & $0.907$ \\
       $X_\mathrm{C}$ (mass fr.)     &   $0.020^{+0.010}_{-0.005}$ & $< 10^{-3}$               & $< 5 \cdot 10^{-4}$               & $3^{+2}_{-2} \cdot 10^{-4}$ & $0.05^{+0.03}_{-0.03}$ \\
       $X_\mathrm{N}$ (mass fr.)     &   $0.004^{+0.004}_{-0.003}$ & $0.008^{+0.004}_{-0.004}$ & $0.008^{+0.004}_{-0.004}$ &  $0.008^{+0.004}_{-0.004}$ & $0.020^{+0.005}_{-0.010}$ \\
       $X_\mathrm{O}$ (mass fr.)     &   $0.005^{+0.004}_{-0.002}$ & $< 10^{-4}$               & $< 10^{-5}$               & $6^{+3}_{-3} \cdot 10^{-5}$ & $0.015^{+0.002}_{-0.002}$ \\
       $X_\mathrm{Ne}$ (mass fr.)    &   $0.005^{+0.003}_{-0.003}$ & $0.010^{+0.005}_{-0.005}$ & $0.002^{+0.002}_{-0.002}$ & $0.003^{+0.002}_{-0.002}$ & ($0.01$) \\
         $X_\mathrm{Fe}$ (mass fr.)    &   $4.8\cdot 10^{-4}$  & $1.0\cdot 10^{-3}$   &  $8\cdot 10^{-4}$    &  $8\cdot 10^{-4}$     & $0.8\cdot 10^{-3}$ \\
         $L_\mathrm{mech}$\,[erg\,s$^{-1}$] &  $6.4 \cdot 10^{36}$ & $8.2 \cdot 10^{36}$ & $2.6 \cdot 10^{36}$ &  $3.4 \cdot 10^{36}$ & $1.1 \cdot 10^{38}$ \\
      \hline       
    \end{tabular}
\end{table*}

From our
hydrodynamically-consistent atmosphere models, we infer the effective temperatures of our objects beneath the expanding layers $T_\mathrm{eff}(R_\mathrm{crit})$ as shown in the HRD (Fig.\,\ref{fig:3}). Both the WN/WO as well as the WN2 stars have $T_\mathrm{eff}(R_\mathrm{crit})$ above $130\,$kK, in line with the expected range of envelope-stripped stars during central He-burning \cite{Langer1989,Grassitelli+2018,Graefener+2011}. This is in line with the absence of hydrogen in the spectra of our sample stars. According to our atmosphere models, at most 10\% of hydrogen could be hidden without leaving noticeable imprints on the observed spectra. Yet, as we discuss in detail in the supplementary material, major current single and binary evolution models \cite{Eldridge+2017,Eggenberger+2021} fail to reproduce the HRD positions of our sample stars, with the possible exception of M33WR\,99-1. The models also have problems in reproducing the abundance patterns of the transition stars, which provides evidence for an insufficient treatment of overshooting during core-He burning. In the stages beyond central hydrogen burning, Overshooting, i.e., mixing of material beyond the convective core, is often treated only rudimentary or even disregarded in massive star evolution models. In contrast, our analysis clearly shows the mixing of material which has undergone central He burning with the above layer containing the enriched nitrogen from the CNO cycle \cite{Langer1991}. As the lifetime of such a mixed region at the surface scales with the assumed degree of overshooting, our newly discovered WN/WO stars as well as the known WN/WC stars could provide crucial benchmarks for future stellar evolution models. Notably, this abundance diagnostic for the mixing should not be affected by pending uncertainties of the $^{12}$C$(\alpha,\gamma)^{16}$O reaction rate as the observed C and O surface abundances in this transition stage stem from the early time of core-He burning where adjustments to the $^{12}$C$(\alpha,\gamma)^{16}$O rate have negligible effects \cite{Aadland+2022}.

To reproduce the obtained HRD positions of the WN2 and WN/WO stars, the stars need to have undergone significant prior mass loss, either intrinsically or due to interaction with a companion star. Yet, the current mass-loss rates of the stars are rather moderate. As they only depend on the involved physics and basic stellar properties in the present stage, we use this observation to conclude that our target objects are connected with lower metallicity, as outlined in the method section. 
Interestingly, the low measured mass-loss rates are in sharp contrast to all the explored predictions from current evolution models. The few single and binary models that reach the HRD positions of the WN/WO and WN2 stars\footnote{except for M33WR\,99-1 in the case of recent GENEC models \cite{Eggenberger+2021}}, which we discuss in detail in the methods part, assume the present mass-loss rate to be at least one order of magnitude higher than the observed values, which marks a major insufficiency in current stellar evolution modelling. This has significant consequences for the subsequent stellar evolution as well as the predicted kinetic and ionizing feedback. In stellar population synthesis models, such as Starburst99, invoking older sets of stellar evolution models with higher mass-loss rates \cite{Meynet+1994} is a popular method to reach the WR stage at lower metallicity \cite{Garner+2025,Gunawardhana+2025}. Yet, these models produce WR stars that totally differ in their balance of kinetic and ionizing feedback compared to our observed WN2 and WN/WO stars as we further discuss below.

From our spectral analysis and evolutionary model comparisons, we conclude that the WN2 stars represent hydrogen-free WR stars in a sub-solar metallicity environment with the WN/WO stage being their advanced form. In this regime, the combined effect of luminosity, mass, and chemical composition of the atmosphere leads to relatively weak winds compared to other WR subtypes and yields hot, compact stars that transition to WN2 and then further to WN2/WO1, skipping the WC spectral stage. While rare in nearby galaxies, such objects are expected to be more common at (even) lower metallicities where WR-winds are generally expected to be weak \cite{VinkdeKoter2005,SanderVink2020}. From the studied sample, the inferred current masses of the stars in our sample are all high enough to end up as black holes, presumably via ``direct'' collapse according to most explosion predictions \cite{Woosley+2020,Maltsev+2025}.

\begin{figure}
    \centering
\includegraphics[width=0.99\columnwidth]{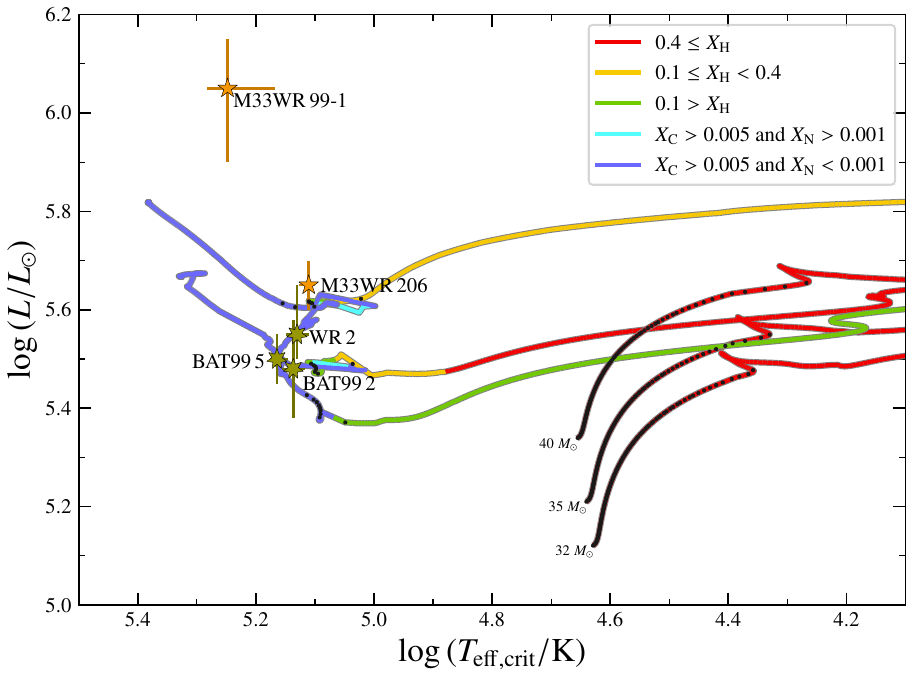}
  \caption{Hertzsprung Russell Diagram (HRD) showing newly calculated GENEC tracks with a custom mass-loss scheme to match the observed mass-loss rates and HRD positions of the analyzed WN/WO (orange) and WN2 (olive) stars. Phases with different surface abundances are highlighted by different coloring of the tracks. The dots along each track mark time steps of 50\,kyr. \label{fig:4}}
\end{figure}

Given that none of the WN2 and WN/WO stars show evidence of a luminous companion, they pose a similar challenge to massive star evolution as most of the hydrogen-containing WN3 stars in the SMC, which are also characterized by weaker winds and lack any signs of a massive companion \cite{Hainich+2015,Schootemeijer+2024}. 
To demonstrate a possible scenario, we present a set of custom GENEC evolution models where we
increase the mass loss in the cool stage to achieve the necessary prior mass removal and use a recent WR mass-loss description that depends on temperature and $L/M$ \cite{Sander+2023}. As evident in Fig.\,\ref{fig:4} and further discussed in the methods part, stars undergoing such a mass-loss scheme spend most of their He-burning lifetime precisely in the region of the observed WN2 and WN/WO stars while also showing the comparably low observed mass-loss rates in this stage without further customizing the WR description. Similar to the WN3 stars in the SMC, the weak emission lines of the WN2 and WN/WO stars -- compared to most other WR spectral types -- will make it hard to detect these stars in unresolved populations.

A transition from a WN-like stage to the WO stage due to the continued loss of the stellar envelope has previously been predicted for stars evolving chemically-homogeneously \cite{Kubatova2019,Szecsi+2025}, assuming their winds are strong enough to appear as WR spectral type. For the first time, we now identify WN/WO-transition objects. The derived chemical abundances of the WN/WO stars clearly exclude a fully homogeneous stellar chemistry at the current point. The necessary mixing of layers undergoing different burning stages also limits potential previous stages of full homogeneity for our sample stars. While reproducing the observed abundance mix of the WN/WO stars for a reasonable amount of time is possible with the current GENEC setup, we use an adapted MESA calculation to demonstrate that enhanced mixing above the He core changes the significantly enhances the lifetime of the transition stage. While further follow-up evolution studies will be required to better constrain the mixing and explain the strong pre-WR mass loss, our demonstration models already indicate the fundamental consequences for massive star evolution, which will also have significant consequences for population synthesis and galaxy evolution models.

\section{Ionizing fluxes and cosmic keystones}

Even when only considering established evolution models, i.e., those not including our studied WN/WO and WN2 stars, massive stars with initial masses above $30\,M_\odot$ contribute most of the ionizing flux budget in a stellar population \cite{EldridgeStanway2022}.
Our studied WN/WO and WN2 stars further strengthen this contribution.
With their comparably thin winds with respect to later-type WR stars, also harder ionizing radiation can escape, making both the WN2 and WN/WO stars important sources of photons below the \ion{He}{ii} ionization edge at $227.85\,$\AA. Due to the direct transition from the WN to the WO stage, they will keep this role throughout their remaining core-He burning lifetime and, as our adapted evolution models indicate, also beyond. As such, they mark interesting candidates to explain low-metallicity star-forming galaxies that do show both narrow and broad \ion{He}{ii} emission but lack signatures of WC stars such as Mrk 1315, SBS 0948+532, or MACS J1149-WR1 \cite{ShiraziBrinchmann2012,Morishita+2025}.

\begin{figure}
    \centering
\includegraphics[width=0.85\columnwidth]{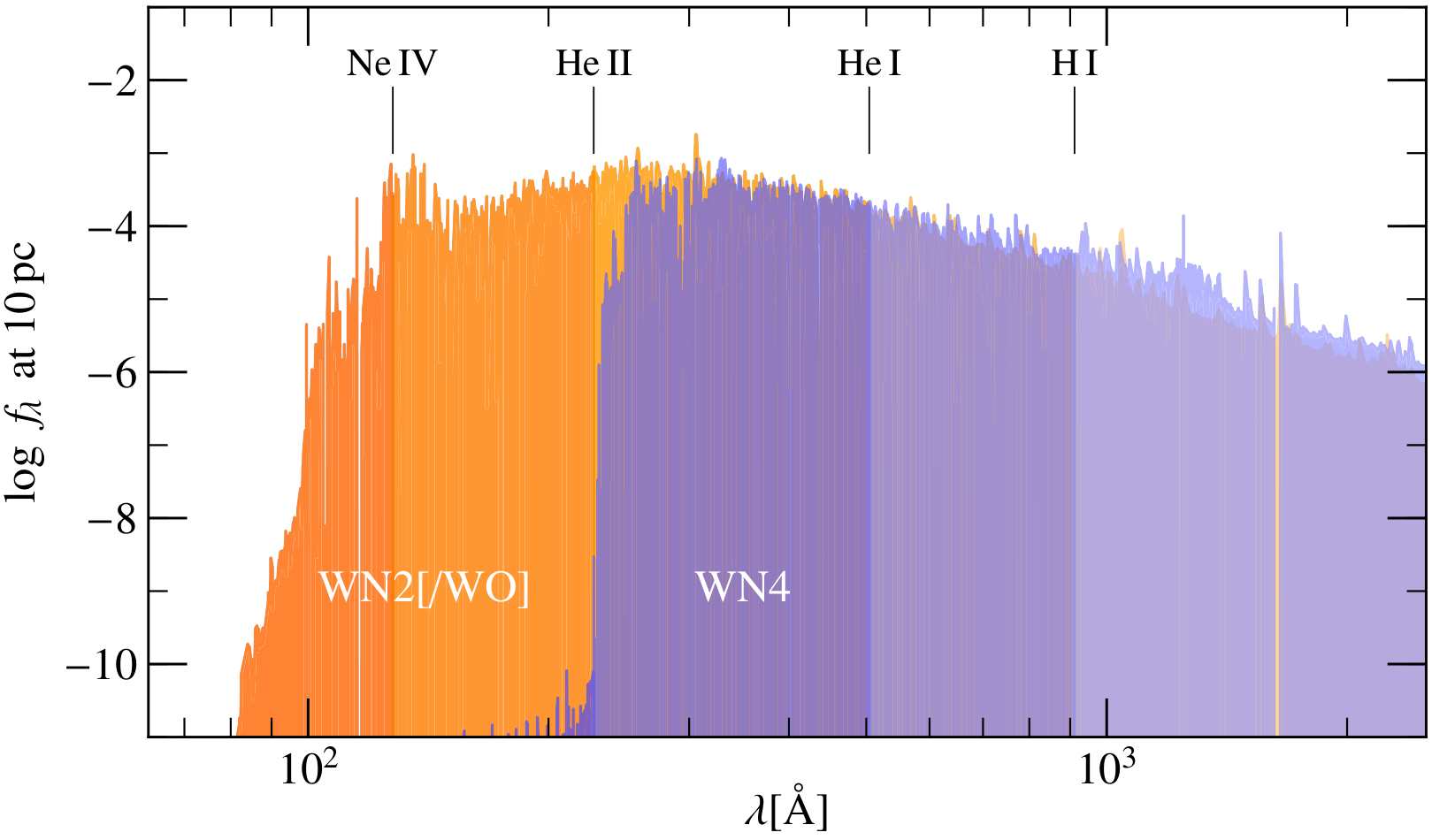}
  \caption{Spectral energy distributions (SEDs) of the newly discovered WN/WO star M33\,WR206 (orange shadings) compared to the typical WN4 star WR\,6 (blue shadings). The stronger winds do not allow a significant photon flux below the \ion{He}{ii} ionization edge at $227\,$\AA, while all the WN2 and WN2/WO stars are strong sources of hard ionizing flux as evident from Table\,2. \label{fig:5}}
\end{figure}

In Table \ref{tab:ionflux}, we list the resulting number of ionizing photons
\begin{equation}
  Q_\mathrm{ion} = 4 \pi R_\ast^2 \int\limits_{\nu_\mathrm{ion}}^{\infty} \frac{F_\nu}{h \nu} \mathrm{d}\nu 
\end{equation}
below several ionization edges for all five studied targets. The \ion{He}{ii} flux of the WN2 stars is approximately equal to those of an $80\,M_\odot$ metal-free Pop III star on the zero-age main sequence \cite{Schaerer2002}, although its current and likely also initial mass are much lower. 
As illustrated in Fig.\,\ref{fig:5}, our studied sources do even contribute substantially below the \ion{Ne}{iv} ionization edge at $\lambda(\nu_{\mathrm{Ne}\,\textsc{iv}}) = 127.67\,$\AA\ (cf.\ Table~\ref{tab:ionflux}). WC stars and later-type WN stars do not contribute any notable flux at these short wavelengths. The uncertainties in determining some of the stellar and wind parameters are not resulting in particularly large error bars for the derived ionizing fluxes, though the influence is stronger on the hardest fluxes as indicated by the errors in Table~\ref{tab:ionflux}.
Our models further reveal that even the extremely hot WN2 and WN/WO stars do not have the capability to provide  significant photon fluxes below the \ion{Ne}{v} edge, which is usually associated with AGN ionization activity \cite{Genzel+1998,Cleri+2023}. 
However, the emergent flux in this wavelength regime is a strong function of metallicity \cite{Smith+2002,MartinsPalacios2021} with our test calculations predicting a significant contribution below the \ion{Ne}{v} edge for metallicities $\lesssim$$5\%$ solar. Consequently, counterparts to our objects at earlier Cosmic times with extremely low metallicities are expected to have very hard ionizing radiation fields.

\begin{table*}
   \caption{Decadic logarithm of the number of ionizing photons per second below a given wavelength for the analyzed WN2/WO1 and WN2 stars\label{tab:ionflux}.}
   \centering
   \renewcommand{\arraystretch}{1.3}
   \begin{tabular}{lcccccc|cc}
      \hline 
        Ion           &   Wave- &   WN2/WO1    &      WN2    &      WN2    &  WN2        &   WN2/WO1   &     WN4       &    WC4    \\ 
        Edge          &  length &  {\small\!\!M33WR\,206\!\!}  &  {\small WR\,2}    &   {\small BAT99\,2}  &  {\small BAT99\,5}   & {\small\!\!M31WR\,99-1\!\!} & {\small WR\,6}   & {\small BAT99\,52}  \\ 
                     &    [\AA]   &   [s$^{-1}$] &  [s$^{-1}$] &  [s$^{-1}$] &  [s$^{-1}$] & [s$^{-1}$]  &   [s$^{-1}$] &   [s$^{-1}$] \\ 
      \hline
      \ion{H}{i}     &  $911.55$  &  $49.5^{+0.1}_{-0.1}$  &  $49.4^{+0.1}_{-0.1}$  &  $49.3^{+0.1}_{-0.1}$ & $49.3^{+0.1}_{-0.1}$ &    $49.8^{+0.1}_{-0.2}$   &  $49.3$  &   $49.6$ \\
      \ion{He}{i}    &  $504.30$  &  $49.3^{+0.1}_{-0.1}$  &  $49.2^{+0.1}_{-0.1}$  &  $49.1^{+0.1}_{-0.1}$ & $49.2^{+0.1}_{-0.1}$ &	  $49.7^{+0.1}_{-0.2}$   &  $49.1$  &   $49.2$ \\
      \ion{He}{ii}   &  $227.85$  &  $48.4^{+0.1}_{-0.2}$  &  $48.4^{+0.1}_{-0.2}$  &  $48.3^{+0.1}_{-0.2}$ & $48.4^{+0.1}_{-0.2}$ &	  $49.0^{+0.2}_{-0.3}$   &  $40.9$  &   $41.6$ \\
      \ion{Ne}{iii}  &  $195.49$  &  $48.0^{+0.1}_{-0.2}$  &  $48.2^{+0.1}_{-0.2}$  &  $48.1^{+0.1}_{-0.2}$ & $48.1^{+0.1}_{-0.2}$ &	  $48.9^{+0.4}_{-0.3}$   &  $40.2$  &   $39.9$ \\
      \ion{Ne}{iv}   &  $127.67$  &  $46.9^{+0.2}_{-0.3}$  &  $47.2^{+0.2}_{-0.3}$  &  $47.2^{+0.3}_{-0.3}$ &  $47.2^{+0.2}_{-0.3}$ &	  $48.1^{+0.3}_{-0.3}$   &  $34.4$  &   $33.2$ \\
      \ion{Ne}{v}    & $\,\,\,98.21$ & $42.5^{+0.3}_{-0.3}$ & $44.5^{+0.3}_{-0.3}$  &  $44.5^{+0.3}_{-0.3}$ & $44.2^{+0.3}_{-0.3}$ & $44.9^{+0.3}_{-0.3}$   &  $24.2$  &   $20.8$ \\
      \hline       
    \end{tabular}
\end{table*}

WN2 stars as well as our newly classified WN/WO stars have exceptionally strong influence on their galactic environments as they mark a sweet-spot between strong kinetic and ionizing feedback. While their mass-loss rates are not as high as those of O2 or later-type WR stars with similar luminosities, their mechanical luminosity is comparable to those of highly influential O stars, such as the O3 stars in the Galactic Cyg OB2 association or the LMC R136 region \cite{Doran+2013}. At subsolar metallicity, the WN2, WN/WO, and later WO stages stars can easily outweigh the influence of hundreds of massive OB stars \cite{Ramachandran+2019}. For example, the OB populations of N206 \cite{Ramachandran+2018} in the LMC or the (southern) supergiant shell in the SMC \cite{Ramachandran+2019} produce a combined mechanical luminosity that is at least an order of magnitude lower than what any of our studied targets provide.

Strong ionizing flux beyond the \ion{He}{ii} edge can also be observed via nebular emission. In fact, the optical spectra of both WN/WO stars show clear imprints of nebular He\,\textsc{ii} 4686\,\AA\ as well as other highly-ionized lines. Hot WR stars, such as the objects presented in this work, could explain nebular \ion{He}{ii} emission, but many He\,\textsc{ii} detections in star-forming galaxies also lack the corresponding imprints of broad WR emission lines in their integrated spectra, in particular at lower metallicity \cite{Guseva+2000,ShiraziBrinchmann2012,Kehrig+2018}. Objects like the WN/WO star {M31WR\,99-1} are very hard to detect due to their comparably weak emission lines \cite{NeugentMassey2023}. These objects could easily be hidden in integrated populations and may at least partially resolve the puzzle of the apparent absence of WR features in many regions with nebular He\,\textsc{ii}  \cite{ShiraziBrinchmann2012,Mayya+2023}, especially at significantly subsolar metallicities where the WR-features are known to be weak.
Recently, multiple high-redshift galaxies have been found where the observed ionization cannot be explained by current stellar evolution models \cite{Strom+2017,Sanders+2020,Tang+2023}. Fitting the observations with stellar sources requires additional types of stars with $T_\text{eff} \gtrsim 80\,$kK \cite[e.g.,][]{Cameron+2024,Cullen+2025,Lecroq+2025,Topping+2025}. Hot WR stars with rather weak winds such as our WN2 and WN/WO stars could provide a natural explanation for these observations without requiring to invoke population III stars. While the precise pathway towards the formation of the WN2 and WN/WO stars is still uncertain, our analyzed stars do not require very high initial masses to form, while still being as or more efficient than the generic templates used in the high-redshift galaxy studies.%

The direct WN-to-WO transition at low metallicity reveals a new evolutionary path occupied by stars that produce high fluxes of hard ionizing photons. The level of convective boundary mixing during He burning required to produce a longer-lived transition stage has implications for the final structure of massive stars, affecting the resulting oxygen core mass and thereby the compactness and properties of any resulting black hole remnants. Emission lines from these WR stars are depressed due to their weak stellar winds which complicates their detection. However, the WN/WO transition is likely to be a common evolutionary path for massive stars with low metallicities, producing stellar populations that are important for understanding high redshift galaxies with high ionization emission lines. New generations of stellar evolution and population synthesis models that account for our discovered weak-winded WR stars and the direct WN/WO transition are necessary to interpret properties of high-redshift galaxies and connected phenomena that occur at low metallicities across cosmic time.

\clearpage

\begin{figure}
    \centering
\includegraphics[width=\columnwidth]{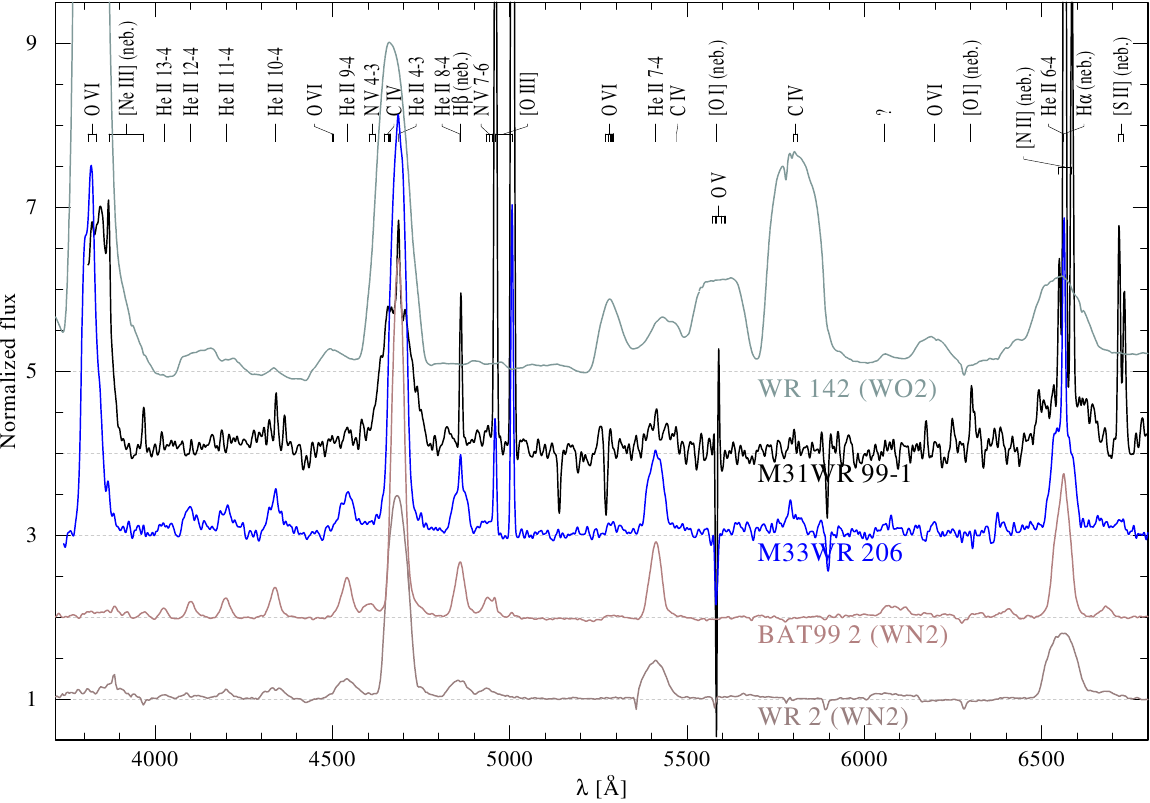}
  \caption{Normalized optical spectrum of the WN/WO stars {M33WR\,206} (blue) and {M31WR\,99-1} (black), compared to two other WN2 stars ({WR\,2}, {BAT99\,2}) as well as the WO2 star {WR\,142}. The normalized spectra have been shifted along the ordinate for clearer distinction between the different objects. \label{fig:2}}
\end{figure}

\clearpage
\renewcommand{\thefigure}{EDF\,\arabic{figure}}
\setcounter{figure}{0}
\renewcommand{\thetable}{EDT\,\arabic{table}}
\setcounter{table}{0}
\section*{Methods}

\subsection*{Classification and naming of the targets}

The star [NM2011] J013510.27+304522.9 (aka M33WR\,206) was discovered by \cite{NeugentMassey2011} and is located in the outskirts of the M33 galaxy. It was initially classified as a peculiar WN2. The WN2 classification was introduced by \cite{vanderHucht+1981} as an extension of the system introduced by \cite{Smith1968}. The WN2 subclass was originally populated only by the Galactic {WR\,2}, but has since been populated further by additions from other galaxies, such as the LMC WR stars {BAT99\,2} and {BAT99\,5}. M33WR\,206 stands out from the other WN2 stars by its strong oxygen lines both at UV and optical wavelengths.

X004418.10+411850.8 (aka M31WR\,99-1) was recently discovered by \cite{NeugentMassey2023} and is located in the Population I ring of M31. With its prominent \ion{O}{vi} 3811\,\AA\ emission, \cite{NeugentMassey2023} classified {M31WR\,99-1} as a WO star, the first of its kind found in M31. Beside \ion{O}{vi}, the second prominent feature is the broad emission around 4686\,\AA, which they interpret as a blend of \ion{He}{ii}\,4686\,\AA\ and \ion{C}{iv}\,4650\,\AA. While this would indeed be typical for early-type WO stars, a comparison of the available optical spectrum with our other targets presented in this work (cf.\ Fig.\,2) shows that the emission peak is more or less symmetric around 4686\,\AA, which raises the question about the \ion{C}{iv} contribution to this line.\footnote{The blend of \ion{C}{iv}\,4650\,\AA\ and \ion{He}{ii}\,4686\,\AA\ can also appear symmetric if both contributions are present, as e.g. seen in \cite{Aadland+2022}, but for typical WC-like carbon abundances the central peak is shifted to shorter wavelengths. This is not the case in M31WR\,99-1, as underlined by the nebular \ion{He}{ii}\,4686\,\AA\ emission sitting precisely on top of the peak from the broad component.} Moreover, there is only a weak emission at \ion{C}{iv}\,5808\,\AA, contrary to the much stronger appearance in all known ``traditional'' massive WO stars \cite{Tramper+2015}. Thus, the object is either so hot that carbon is almost completely in higher ionization stages than \ion{C}{iv} or the carbon mass fraction is considerably lower than in other WO stars. 
Our atmosphere modelling efforts (see below) reveal a small carbon contribution to the blend around \ion{He}{ii}\,4686\,\AA\ but also a noticeable \ion{N}{v} $\lambda\lambda 4933,4944$. With the nitrogen abundance being likely larger than the carbon abundance (see Table~\ref{tab:results}), M31WR\,99-1 is more WN2- than WC-like and we classify it as another WN/WO-transition object. 

A revisit of the subtype classification for M33WR\,206 and M31WR\,99-1 following the schemes for WN and WO stars \cite{Smith+1996,Crowther+1998} reveals that both targets adhere to the WN2 and WO1 types simultaneously, albeit with some caveats. Both WN/WO stars show a weak but detectable presence of the \ion{C}{iv}\,5808\,\AA\ line, which is absent in other WN2 stars. On the other hand, both targets lack the diagnostic \ion{O}{v}\,5590\,\AA\ and \ion{O}{vii}\,5670\,\AA\ lines usually present in WO1 star spectra, e.g., in the [WO1] star NGC 6905 \cite{Gomez-Gonzalez+2022}. As the spectra lack any of the characteristics of later WN and WO subclasses, we therefore classify both M33WR\,206 and M31WR\,99-1 as WN2/WO1. This marks the first (modern) assignment of the WO1 subclass among massive WR stars, which was so far only populated by low-mass central stars of Planetary Nebulae.

Following a similar scheme as the Galactic WR notation \cite{RossloweCrowther2015}, we assign WR number designations for [NM2011] J013510.27+304522.9 and X004418.10+411850.8, namely {M33WR\,206} and {M31WR\,99-1}, respectively. To obtain the numbers, we use the RA-ordered list of WR stars taken from \cite{Neugent+2012} for M31 and \cite{NeugentMassey2011} for M33. For M31, this scheme was first adopted in \cite{Sander+2014} with later WR discoveries now added in a simple numbered order, such as the transition-type star {M31WR\,84-1} \cite{Shara+2016} or now X004418.10+411850.8 \cite{NeugentMassey2023} getting the number {M31WR\,99-1}.

\subsection*{Observational data}
\label{sec:obsdata}

We obtained the first ever UV spectra of {M33WR\,206} with the Cosmic Origin Spectrograph \cite{Green+2012} aboard the Hubble Space Telescope (HST, proposal GO 16170, PI A.A.C.\ Sander). Given the large distance, we used the low-resolution, gapless G140L/800 setting \cite{Redwine+2016} covering wavelengths up to $1950\,$\AA. The isolated nature of M33\,WR206 \cite{NeugentMassey2011} is confirmed by the target acquisition image. For the optical spectrum of {M33WR\,206}, we make use of the data provided by \cite{NeugentMassey2011}. These spectra were taken with the 6.5m MMT in Arizona using the Hectospec spectrometer and cover a wavelength range from 3700 to 7500\,\AA\footnote{The raw data further covers the range from 7500 to 9000\,\AA, which we exclude due its contamination with second-order blue light.}.
To recover the spectral energy distribution, we further use UBVRI photometry \cite{Massey+2006}, UV photometry from GALEX \cite{MuddStanek2015}, and optical Pan-STARRS1 photometry \cite{Chambers+2016}. The source is not detected at longer wavelengths.

For {X004418.10+411850.8} aka {M31WR\,99-1}, we use the public spectra \cite{NeugentMassey2023} taken with Binospec on the $6.5\,$m MMT as well as ``Panchromatic Hubble Andromeda Treasury'' (PHAT) photometry \cite{Williams+2014}. Only this set of HST photometry is precise enough to resolve {M31WR\,99-1}. The two blue-most photometry values are flagged as poor quality, but their inferred fluxes align well with the spectral energy distribution obtained from the photometry at longer wavelengths. 

For the known WN2 star WR\,2, we use optical spectra from \cite{Hamann+1993} as well as archival IUE (MLFPM, PI Massey), FUSE (G039, PI Herald) and SPITZER \cite{Ardila+2010} spectra. Beside Gaia DR3 \cite{GaiaDR3} and 2MASS \cite{Cutri+2003} photometry, we also consider the narrow-band photometry for WR\,2 from \cite{Massey1984}. For the analysis and discussion of the LMC WN2 stars {BAT99\,2} and {BAT99\,5} we employ optical spectra \cite{Foellmi+2003} as well as IUE archival spectra (PI024, PI Bianchi) for BAT99\,2 and brand-new COS G140L/800 spectra for BAT99\,5 (HST, proposal GO 17426, PI A.A.C.\ Sander). Supporting photometry to determine the SED and reddening uses near-UV data from XMM-OM \cite{Page+2012} (for BAT99\,2 only), narrow-band $v$ photometry \cite{Breysacher+1999}, Gaia DR3, JHK \cite{Bonanos+2009}, and Spitzer \cite{SSTSL2} data. The WO2 comparison in Fig.\,1 shows the reduced HST data of {LMC195-1} by \cite{Aadland+2022} while the comparison in Fig.\,2 shows the X-Shooter spectra from the analysis of {WR\,142} in \cite{Tramper+2015}.

\subsection*{WN2 sample}

To draw our conclusions on the WN to WO transition and its connection to the WN2 class, we analyzed all known WN2 stars that had available spectra of sufficient quality and were not significantly diluted by a companion. Among the 709 massive WR stars known in the Milky Way\footnote{obtained from the Galactic WR catalogue, v1.31, last updated in Feb.\,2025, online available at \url{http://pacrowther.staff.shef.ac.uk/WRcat/}}, only one star (WR\,2) is of the WN2 subtype. Among the 154 LMC WR stars \cite{Neugent+2018}, two WN2 stars exist, namely BAT99\,2 and BAT\,5. These are analyzed together with WR\,2 and the two WN/WO stars in detail in this work. Among the only 12 known WR stars in the SMC, none are of WN2 subtype. 

In M31, 173 WR stars are known \cite{Neugent+2012,NeugentMassey2023}. Beside the analyzed WN/WO star, there is one more star classified as a peculiar WN2, namely LGGS J004455.82+412919.2 (M31WR\,125) \cite{Neugent+2012}. While formally correct with respect to the subtype-defining lines, the WN2 classification for this object is a bit misleading as there is a clear \ion{N}{iv}\,7110\,\AA\ feature as well as indications of \ion{N}{iii}\,4634,42\,\AA\ which might have been missed due to a different normalizing of the observation. The star also shows a strong \ion{C}{iv}\,$\lambda\lambda 5801,5012$ emission, which is untypical for the WN2 class. This object also cannot be a WN/WO star as not only the stellar \ion{O}{vi} $\lambda\lambda 3811,3834$ is lacking, but also the nebular emission lines in the spectrum point to a lower ionization stage. Most likely, M31WR\,125 is not a single object, but a source where one or more WR stars with cooler effective temperatures than our studied sample stars are diluted by other OB stars.

With IC\,1613 containing only one WR star (of type WO) and NGC\,6822 as well as IC\,10 containing WR stars poorly covered in terms of suitable spectroscopic observations (with none of the WRs classified so far being of WN2 type), this leaves M33 with its significant metallicity gradient. Beside our analyzed WN/WO transition star (M33WR 206), there are five stars classified as WN2 out of the 206 known WR stars in M33 \cite{NeugentMassey2011}. Four of these WN2 stars, namely LGGS J013233.24+302652.3 (M33WR\,3), LGGS J013245.74+303854.4 (M33WR\,6), LGGS J013307.50+304258.5 (M33WR\,16), and LGGS J013359.79+305150.0 (M33WR\,54) clearly show absorption lines from a companion in their optical spectra. While the fifth object (LGSS J013232.07+303522.4  aka M33WR\,1) does not show clear absorption-line features, our obtained HST UV imaging clearly indicates that at least three targets are entering the UV spectrum with most of them likely also affecting the available optical spectrum. While analysing these multi-component sources would be interesting in itself, neither the available data is sufficient for that, nor is it likely that the WR components in most of these targets are actually of WN2 subtype. As the WN2 spectral type is defined by the absence of lower nitrogen ions, this classification is formally also obtained when lower-ionization nitrogen lines get diluted by the companion in the case of low resolution and low signal-to-noise. Therefore, none of the other M33 WN2 targets are suitable for the current study beside the WN/WO transition-star M33\,WR206.

\subsection*{On the possibility of a WN+WO binary or chance alignment}

The simultaneous presence of WN- and WO-type features in the spectra of {M33WR\,206} and {M31WR\,99-1} raises the question whether they could be explained as composite spectra of binary stars, 
or as accidental superposition  of physically unrelated stars in the line of sight. Observing a binary WN+WO star 
is very unlikely since the WR phase is short \cite{Crowther2007,Pauli+2022,Josiek+2024} and would require that both stars evolve almost simultaneously. At present, only two massive WR+WR binaries 
with both components being hydrogen-free are known -- WR 48a and WR 70-16 (``Apep'') \cite{Williams2019, Callingham+2020}.  Both these objects are exceptionally (for non-degenerate binaries) bright in X-rays (0.2-10.0\,{\rm keV}) 
with luminosities $L_{\rm X}\propto  10^{35}$\,erg\,s$^{-1}$ explained by wind-wind collisions \cite{Zhekov2011, Callingham+2020}. If {M33WR\,206} and {M31WR\,99-1} had
comparable X-ray luminosities, they would already have been detected  \cite{Stiele2011, Williams2015}. In general, strong X-ray emission of WO and WC stars is a reliable indicator of binarity \cite{Oskinova2003}. On the other hand, putatively single WN and WO stars are X-ray faint  \cite{Oskinova2009, Huenemoerder2015}.
Among our sample stars only WR\,2 is detected in X-rays allowing to determine its X-ray luminosity,  $L_{\rm X}\approx 4\times 10^{32}$\,erg\,s$^{-1}$, which is comparable to other single WN stars \cite{Zhekov2011}. We conclude that X-ray properties of our sample stars are consistent with a single star nature.

\begin{figure*}[htbp]
    \centering
    \includegraphics[width=0.9\textwidth]{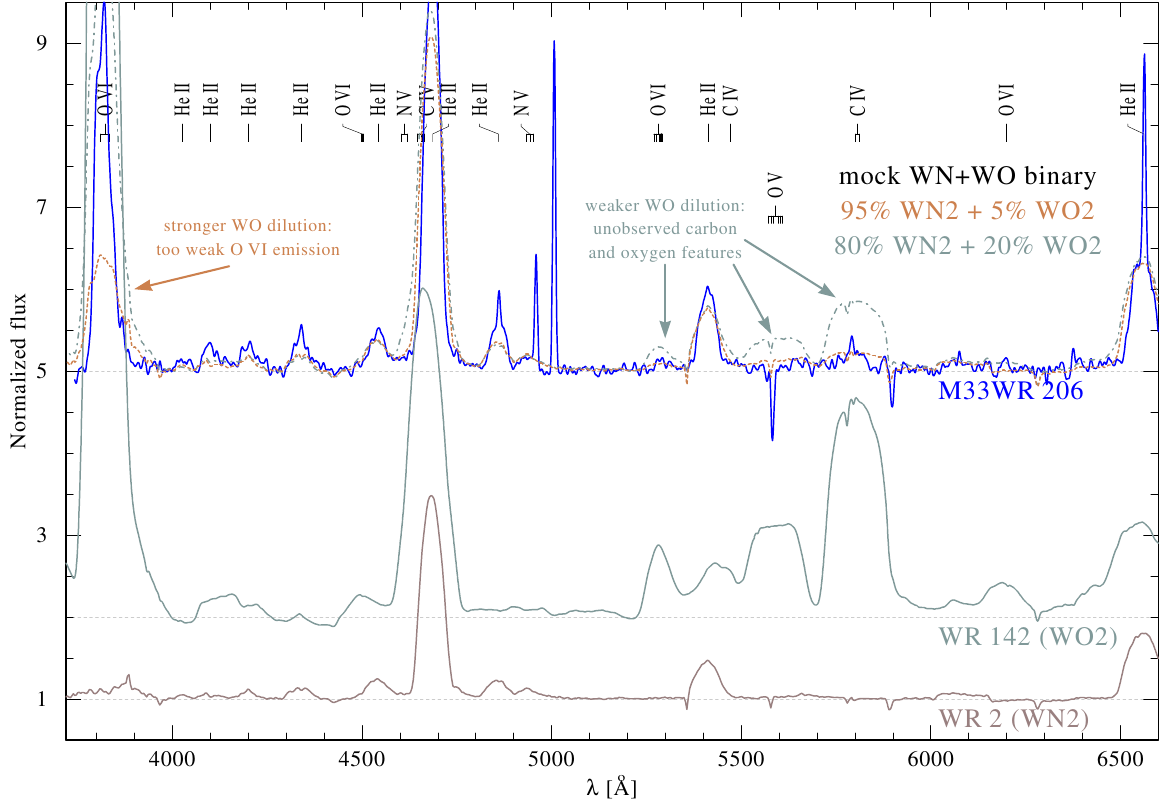}\\[1em]
    \includegraphics[width=0.9\textwidth]{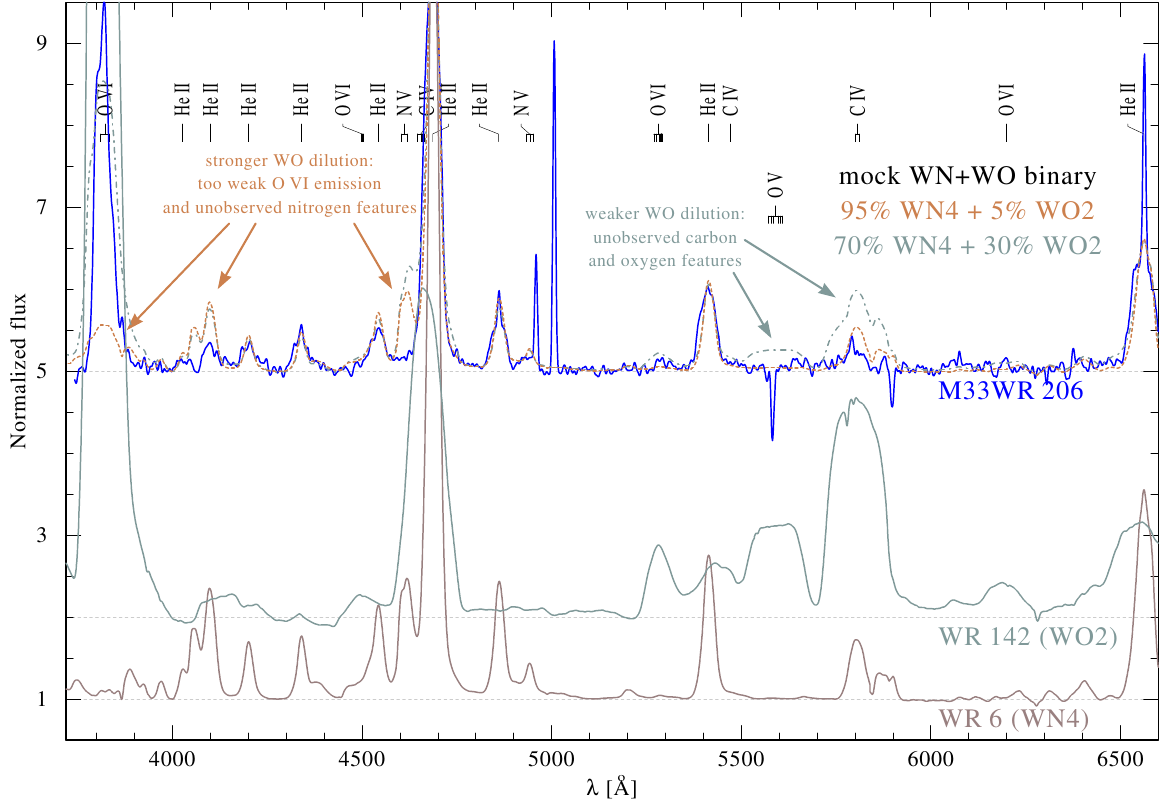}
    \caption{Upper panel: Comparison of the normalized optical spectrum of M33WR\,206 with those of the WN2-star WR\,2 and the WO-star WR\,142. To check whether the unique appearance of M33WR\,206 could also be the product of a WN and WO star, we show two weighted combinations of the latter two stars with the WN2-star providing either 80\% (dash-dotted line) or 95\% (dotted line) of the flux to the optical spectrum. In either case, unobserved features appear with the most prominent ones being highlighted in the figure.\\ 
    Lower panel: Similar comparison, but now using the WN4 star WR\,6 as the WN component, illustrating that for cooler WN subtypes even more additional features would appear.} 
    \label{fig:bincheck}
\end{figure*}

Any chance alignment of a WN and a WO star is also highly unlikely. The consistency of the emission line widths produced by the single-model fit of the spectrum for each of the WN/WO objects would imply that not only the coincidence of a WN and WO star is required, but also their radial velocities do not leave any noticeable impact and their wind velocities would have to be similar, which altogether lowers the chances even further. 
 To investigate the possibility of a chance alignment more quantitatively, we consider whether the spectral appearance of M33WR\,206 could be mimicked by combining typical WN and WO spectra. In the upper panel of Fig.\,\ref{fig:bincheck}, we show an example where we add the spectra of WR\,2 and WR\,142, prototypes for WN2 and WO2-type stars. While these are stars in the Milky Way, we do not aim to precisely reproduce the line widths and allow a free scaling of the total emission line strengths to resemble the observed WN/WO-appearance when co-adding the specific spectra. As the spectrum of {M33WR\,206} was initially classified as WN2pec \cite{NeugentMassey2011}, it is no surprise that the WN star would need to contribute more than half of the total flux. To sufficiently suppress most of the features present in non-transition WO stars, even a 95\% contribution is necessary (dotted line in Fig.\,\ref{fig:bincheck}). Yet, this would suppress the characteristic O\,VI\,3811\,\AA-line way too much. While this could in principle be countered by assuming a higher oxygen abundance or a stronger wind mass-loss, both of these effects would also increase other oxygen lines in the optical, which we needed to suppress in the first place. This is illustrated in the dash-dotted second addition example in the upper panel of Fig.\,\ref{fig:bincheck}, where the WN2 star contributes ``only'' 80\% in the optical and multiple carbon and oxygen lines are clearly stronger than observed. 
 
 Given the limited amount of spectral features for the observed WN/WO stars, considering any later WN or WO spectral types would yield further unobserved spectral imprints. The lower panel of Fig.\,\ref{fig:bincheck} illustrates this by using a WN4 star (here WR\,6) which immediately yields additional nitrogen lines in the blue optical range. The solution where the WO contribution is increased to match the O\,VI\,3811\,\AA-emission is even worse as the additional unobserved carbon and oxygen lines would now appear together with the nitrogen features.
 While it is hard to completely rule out a crowding effect at such extragalactic distances, our examples illustrate that the superposition of a WN and WO star is highly unlikely from the obtained spectral appearance. While in principle there might be WR model combinations that could yield the combined spectrum, the parameters for each of the two components would have to be very close to what we can reproduce more than sufficiently with a single WR model. Therefore, we consider the WN/WO-star solution far more realistic than any binary or crowding scenario.

\subsection*{Stellar atmosphere modeling}
\label{sec:atmmodel}

To analyze the observations and draw theoretical conclusions, we employ the PoWR model atmosphere code \cite{Graefener+2002,HamannGraefener2003,Sander+2015}. Beside calculating standard models assuming a fixed mass-loss rate $\dot{M}$ and velocity field $v(r)$, we further employ the hydrodynamical branch (PoWR$^\mathrm{\textsc{HD}}$) to get self-consistent wind parameters by coupling hydrodynamics and radiative transfer  \cite{Sander+2017,Sander+2018,Sander+2020}. We further include the recent option to account for non-monotonic regions in the hydrodynamic solution \cite{Sander+2023}.

All models presented in this work use detailed atomic for He, C, N, O, Ne, Na, Mg, Al, Si, P, S, Cl, Ar, K, Ca, and the Fe group. The latter is treated in the ``superlevel approach'' \cite{Graefener+2002} with further splitting of the energy bands into levels of different parity. Table~\ref{tab:datom} contains a list of all considered ions, levels, and their line transitions.

In hydrodynamically-consistent models, the wind parameters are fundamentally tied to the stellar parameters, implying that different mass-loss rates $\dot{M}$ and terminal velocities $v_\infty$ can only be obtained when at least one of the stellar parameters or the wind clumping description is changed. In this work, we make extensive use of the option to vary the underlying stellar mass $M$ for a fixed mass-loss rate \cite{Sander+2020}. For the same stellar luminosity $L$ and radius $R_\ast$, an increase in $M$ will yield a decrease in $\dot{M}$ and an increase in $v_\infty$. The terminal velocity further depends  on the assumption of the maximum clumping factor $D_\infty$ with higher values of $D_\infty$ yielding a higher $v_\infty$ for an otherwise fixed set of input parameters.

All targets were treated as single stars in the atmosphere analysis due to the absence of clear signs of multiplicity in the available spectra. For {WR\,2}, a close-by B2.5 main sequence star was discovered a few years ago using high-resolution GMOS spectroscopy \cite{Chene+2019}. Yet, the imprint of the small lines of the companion is rather weak and the star contributes less than 10\% of the total flux. Hence, this is likely a background object not physically bound to the WR star \cite{Chene+2019}. In our study, we refrain from an explicit two-component spectral analysis for {WR\,2} as there is no evidence for significant dilution of the emission lines and the star was further classified as a single star in detailed radial velocity investigations \cite{Dsilva+2022}.

\subsection*{Quantitative spectral analysis of the WN2 and WN/WO stars}

For the quantitative spectral analysis, we compare the above mentioned observations to dedicated sets of hydrodynamically-consistent stellar atmosphere models. To obtain the luminosities and reddening, we reproduce the spectral energy distribution given by the flux-calibrated UV spectra and the photometry. All model fluxes were diluted according to the distance of the target. We employ a distance modulus (DM) of $11.90\,$mag for WR\,2 \cite{Chene+2019}, while all extragalactic targets were calibrated with the distance to the respective galaxy. For the LMC, host of BAT99\,2 and BAT99\,5, we assume a distance of $\mathrm{DM} = 18.48\,$mag \cite{Pietrzynski+2019}, for M31 we used $\mathrm{DM} = 24.38\,$mag \cite{Riess+2012}, and for M33 we employed $\mathrm{DM} = 24.62\,$mag \cite{Breuval+2023}. To obtain the reddening $E_{B-V}$, we mostly employ a standard reddening law \cite{Fitzpatrick1999} with the parameter $R_V = 3.1$, except for WR2, where $R_V = 3.0$ slightly improves the fit. We obtain $E_{B-V}$ values of $0.50$ for WR2, $0.14$ for M33WR\,206, and $1.30$ for M31WR\,99-1. For the LMC stars BAT99\,2 and BAT99\,5, we use a different reddening law \cite{Howarth1983} with $R_V$ fixed to $3.2$. We obtain $E_{b-v} = 0.13$, which corresponds to $E_{B-V} \approx 0.16$, for BAT99\,2 and $E_{b-v} = 0.18$ (i.e., $E_{B-V} \approx 0.22$) for BAT99\,5.

To reproduce the line spectra, the normalized UV and optical spectra are compared to the observations with parameters being adjusted until a suitable reproduction is achieved. As the normalization in the UV regime is non-trivial, we normalize with the continuum calculated by the stellar atmosphere model. Due to the coupling of stellar and wind parameters in the hydrodynamically-consistent models, the parameter degeneracy in the optically-thick wind regime \cite{HamannGraefener2004,Lefever+2023} is broken and usually only a narrow combination of parameters provides a suitable fit. For example, for otherwise fixed parameters, models with a larger (critical) radius will yield narrower and stronger emission lines \cite{Sander+2023} while models with a smaller radius will yield an opposite trend. The first estimate of the wind-strength is mainly set by the helium emission lines as the effects of abundance adjustments in CNO on the mass-loss rate and the terminal velocity are of second order for the objects considered in this study. Given the emission-line dominated spectra, the models were calculated with the ``flexible mass'' option where the stellar mass is varied for a given luminosity and mass-loss rate until a hydrodynamically-consistent solution is reached \cite{Sander+2020}. While the long calculation times do not allow to establish high-dimensional model grids, the uncertainties can sufficiently be estimated by varying the individual model input parameters until the fit quality becomes insufficient. For output-dependent quantities such as the terminal velocity, the ionizing fluxes, or the effective temperature at $\tau = 2/3$, the uncertainties were determined by taking the minimum and maximum values for these quantities among the set of models that had input parameters inside the acceptable uncertainty regime.

\subsection*{Identification of carbon, nitrogen, and neon}

Albeit the original definition of the WN2 subclass implies an absence of nitrogen in the optical spectrum \cite{vanderHucht+1981}, all well-studied WN2 stars actually show at least one visible nitrogen line (\ion{N}{v} $\lambda\lambda 4933,4944$) in their optical spectra when spectra with sufficient S/N are available. This optical \ion{N}{v} line is quite weak and thus easy to miss when inspecting the data by eye. A more clear signature is the \ion{N}{v} $1239,1243$ doublet in the UV, which shows a pronounced P\,Cygni profile in all WN2 stars covered so far ({WR\,2}, {BAT99\,2}, {BAT99\,5}), including {M33WR\,206}. Carbon lines are usually absent in WN2 stars, with neither \ion{C}{iv} $\lambda\lambda 5801,5012$, nor \ion{C}{iv} $\lambda\lambda 1548,1551$ being present. For {BAT99-5}, a very weak feature is present for \ion{C}{iv} $\lambda\lambda 1548,1551$, requiring only a small carbon mass fraction of $3 \cdot 10^{-4}$ to be reproduced. For
{M33WR\,206}, there is instead a clear, strong P\,Cygni profile in \ion{C}{iv} $\lambda\lambda 1548,1551$ and a weak, but notable \ion{C}{iv} $\lambda\lambda 5801,5012$ optical emission line. The differences in the UV are illustrated in Fig.\,1, where the spectrum of {M33WR\,206} is compared to the spectra of the three WN2-type stars, namely {WR\,2} in the Milky Way and {BAT99\,2} as well as {BAT99\,5} in the LMC.

When comparing the optical spectra (Fig.\,2), {M33WR\,206} does not only share similarities with known WN2 stars but also with the recently discovered object  {X004418.10+411850.8}. \cite{NeugentMassey2023}. Initially {X004418.10+411850.8} aka {M31WR\,99-1} was classified as a WO star, but the spectrum lacks the strong carbon features typical for other WO stars (such as the prominent example {WR\,142}). Instead, the striking similarities of {M31WR\,99-1} to the spectral appearance of {M33WR\,206} motivated us to add this object to our analysis.

Based on the discovery of \ion{Ne}{v} in the UV spectrum of our hot, peculiar WN2 target, we revisited other WN2 stars with available UV spectra. Indeed, in both the Galactic WR2 and the LMC WN2 stars BAT99\,2 and BAT99\,5 the feature is visible and not reproduced in the respective earlier analyses \cite{Hamann+2006,Hainich+2014} which did not contain neon. Given the hot derived stellar temperatures, we added neon to the atmosphere models for these three targets. Again, we find that not \ion{N}{iv} $\lambda\lambda 1719$-$1722$ but rather \ion{Ne}{v} $\lambda 1718$ is the cause of the observed line feature. The comparison of the models with available NUV spectra for WR\,2 and BAT99\,2 further indicates that neon is a major contributor to an emission feature around $2250\,$\AA\ which cannot be explained just with a series of He\,II lines. Unfortunately, this region is of poor S/N quality and thus not suitable to draw stronger conclusions. Using all available diagnostics, we derive mass fractions of $\sim$$1\%$ (WR\,2) and $\sim$$0.3\%$ (BAT99-2, BAT99-5).

\subsection*{Transition from WC/WO to WN/WO spectral appearance}

\begin{figure*}[htbp]
    \centering
    \includegraphics[width=0.8\textwidth]{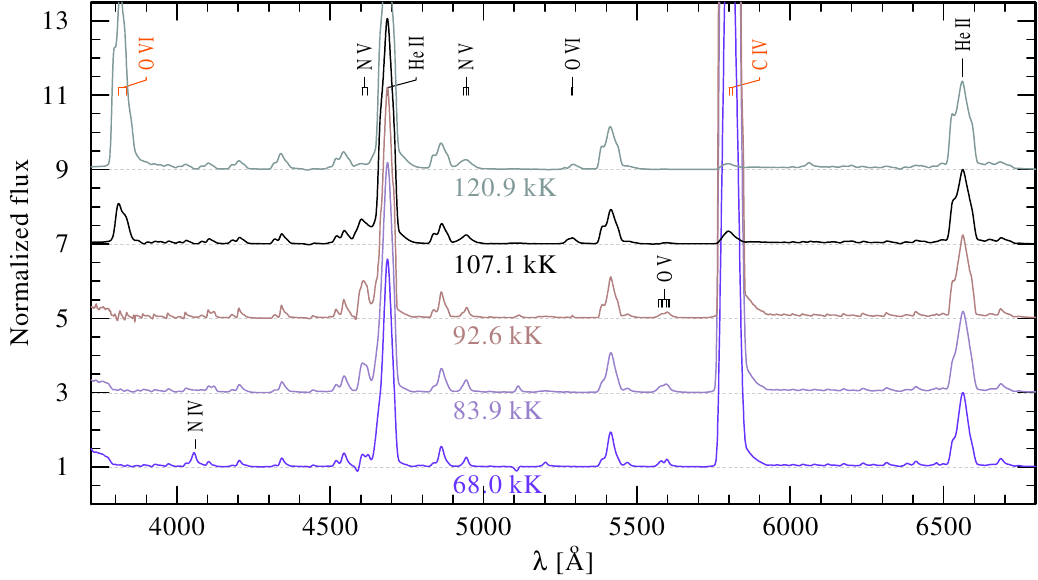}
    \caption{Predicted optical spectra for different atmosphere models including the same luminosity, mass, chemical composition, and wind parameters, but differ in their effective temperatures ($T_{2/3}$). For hotter temperatures, the spectral diagnostics change significantly. Most prominent is the disappearance of  \ion{C}{iv}\,$\lambda\lambda$5801/5012 and the appearance of \ion{O}{vi}\,$\lambda\lambda$\,3811/3834.} 
    \label{fig:wrtempseq}
\end{figure*}

In order to cross-check our interpretation of WN/WO stars marking a hotter counterpart of WN/WC-transition stars, we calculate a set of atmosphere models where all stellar and wind parameters are prescribed. Choosing a chemical composition similar to the WN/WO-star M33WR\,206, we keep all parameters fixed and vary only the effective temperature in this model set. In Fig.\,\ref{fig:wrtempseq}, we show the resulting optical spectra for five models, highlighting prominent spectral emission lines. For high $T_\text{2/3} \gtrsim 100\,$kK, we get the characteristic appearance of the newly discovered stars with a strong \ion{O}{vi} $\lambda\lambda$3811/3832 emission feature and an almost absent \ion{C}{iv} $\lambda\lambda$5802/5812. For lower temperatures, the ionization of the atmosphere changes and \ion{O}{vi} disappears while \ion{C}{iv} $\lambda\lambda$5801/5812 becomes very strong as seen in WN/WC stars (e.g., BAT99 88 in the LMC).  As the strength of these two emission doublets are major criteria for the WC or WO stars, we demonstrate that the WN/WC transition can be replaced by a WN/WO transition if the stars maintain to stay in a regime of higher $T_\text{2/3}$.

\subsection*{Low metallicity as the origin of the weaker WN2 and WN/WO winds}

\noindent%
Our quantitative spectroscopic analysis of the WN/WO stars as well as the WN2 stars shows that these stars have rather low mass-loss rates compared to later-type WR stars and thus the bulk of the currently known WR population (which is naturally biased towards higher-metallicity objects).
From detailed modelling of the driving of WR-type winds \cite{Sander+2020,SanderVink2020,Sander+2023}, we can infer that the wind mass-loss rate increases with higher luminosity-to-mass ($L/M$), higher metallicity ($Z$), or lower $T_\mathrm{eff}(R_\mathrm{crit})$. With the exception of M31WR\,99-1, Fig.\,3 highlights that the difference in $T_\mathrm{eff}(R_\mathrm{crit})$ between strong-winded stars such as WR\,6 and a WN2 star such as BAT99\,2 is relatively small, if not the same within error bars. Consequently, the different appearance of WN/WO and WN2 stars compared to strong-winded stars must either stem from a lower $L/M$ or a lower $Z$. 

Given the relative rareness of WN2 stars compared to later-type hydrogen-free WN stars in the Milky Way and the LMC, a lower $L/M$ solution is very unlikely. As the bulk of the hydrogen-free WN population is most certainly in the stage of central helium burning and the surface abundances of the WN2 and WN/WO stars even suggests that our stars are rather a bit more evolved, there is no known evolutionary scenario that predicts a large, inherent decrease in the $L/M$-ratio from central Helium-burning onwards. Hence, the only way to create an ``underluminous'' He-burning star would be to add mass to its surface, e.g., in a stellar merger. As there is no hydrogen in the spectrum, such a merger would have to be with another He-burning star with any possible hydrogen envelope being lost in the merger process. Still, such a merger should eventually relax to a status where the burning core gets bigger and the $L/M$-ratio becomes more ``normal'' again. This would imply that all of our studied objects are caught in the short pre-relaxation stage.

The more likely explanation is therefore that both the WN2 and the WN/WO stars represent a stage that mainly occurs at lower $Z$, where earlier WR subtypes with usually weaker winds are generally more common $Z$ \cite{Crowther2007}. Clearly, BAT99\,2, BAT99\,5 and M33\,WR206 are sitting in a sub-solar metallicity environment. Even the Galactic WN2 star WR2 is likely located in the region beyond $9\,$kpc from the Galactic Center given its  Galactic coordinates and presumed distance \cite{Chene+2019} and could thus be in a sub-solar iron abundance regime. M31WR\,99-1 provides a challenge as it is located in the star-forming population I ring of M31 between 9 
and 15 kpc galactocentric distance \cite{vandenBergh1964}. While it is impossible to rule out higher-metallicity solutions for this object, measurements of the metallicity gradient in M31 \cite{Sanders+2012,Pena+2019,Bhattacharya+2022} indicate that sub-solar iron abundances can exist in this region. As we discuss further below, the object does not appear ``overmassive'' if it has an LMC-like iron abundance. A higher, potentially still sub-solar metallicity, could also be accommodated if the luminosity of M31WR\,99-1 is overestimated due to the photometry being affected by unresolved companions or crowding.

\subsection*{Mass discrepancy between atmosphere models and structure constraints}

Our hydrodynamically-consistent models provide us with the opportunity to infer the stellar mass. While these masses are not as precise as orbital masses, they can be used as sanity checks and compared to the maximum possible mass given by stellar structure constraints. As the luminosities and wind parameters are constrained by the photometric and spectral data, the resulting mass considerably depends on the assumed abundances, in particular iron as the strongest source of line-driving opacity. Running models with different iron abundances and comparing the resulting masses with structural constraints therefore provide plausibility checks for our conclusion that most (or even all) of the known WN2 and WN/WO stars are the product of a low-metallicity environment.

For hydrogen-free stars, the minimum luminosity-to-mass ratio and thus the maximum mass for a given luminosity is obtained for a chemically homogeneous star. We thus compare the mass derived from the hydrodynamically-consistent atmosphere models with masses obtained from a luminosity-to-mass relation for chemically homogeneous He stars \cite{Graefener+2011}. 
For M31WR\,206, the homogeneous mass for the derived luminosities is $\sim$$19\,M_\odot$, while it is $\sim$$15\,M_\odot$ for BAT99\,2 and BAT99\,5. The model results with $\sim$$18\,M_\odot$, $14\,M_\odot$, and $13\,M_\odot$ are slightly lower, which would not be unusual as our stars will not be chemically homogeneous in their interior, not least as all the studied stars are most-likely further evolved than a star on the helium zero age main sequence.

For WR\,2, our preferred model yields a mass of $17.4\,M_\odot$. Formally being slightly higher than the $16.7\,M_\odot$, this is well within the error margin (see Table\,1) of the derived mass. Yet, we require to consider an iron abundance of about $62\%$ solar as all models with higher iron abundance would only lead to larger masses. For example, a suitable fit assuming solar iron abundance would require a model with $\sim$$24\,M_\odot$. While we cannot rule out such a solution completely, it would require exotic scenarios to explain the underluminous nature of the star. Consequently, the conclusion that the star has an iron abundance of $\sim$$60\%$ solar and thus is a product of a sub-solar metallicity environment is strongly preferred. Models with metallicities below $60\%$ solar are also possible, but did result in a worse spectral fit.

Among the four studied targets, M31WR\,99-1 is the most luminous and thus potentially also the most massive object. For a solar iron abundance, its required stellar mass would be $\sim$$51\,M_\odot$, making it the most massive envelope-stripped star known to date. However, the structure relation for chemically homogeneous stars predict an upper limit of $\sim$$34\,M_\odot$, significantly lower than the obtained value. While again such an underluminous situation could potentially be possible due to an exotic merger, we also investigate subsolar iron solutions. Our final model reproducing the spectrum with a half-solar (i.e., LMC-like) iron composition and a bit a higher effective temperature yields a mass of $33.7\,M_\odot$, in line with the structural constraints.

\subsection*{Consequences of a lower intrinsic luminosity of M31WR 99-1}

With the high luminosity inferred for M31WR 99-1 and the limited signal-to-noise of the available spectra, we cannot rule out that this target is subject to dilution by one or more objects, even if they might not be physically forming a gravitationally bound multiple system. From the available spectral features, we can rule out a WN+WO system which would yield many more emission line features. While there is no immediate signature for absorption-line stars in the spectrum, these could potentially be hidden in the noise and we performed a few tests on how a dilution of the emission lines by a factor two and a reduction of the intrinsic luminosity to $\log L/L_\odot = 5.6$, more similar to the other targets, would impact the resulting model necessary to reproduce the diluted spectrum. In case of dilution, the emission measure and thus the luminosity-scaled mass-loss rate, which can be expressed in the form of the so-called ``transformed'' mass-loss rate
\begin{equation}
  \dot{M}_\mathrm{t} = \dot{M} \sqrt{f_\mathrm{cl}} \frac{1000\,\mathrm{km\,s}^{-1}}{v_\infty} \left(\frac{10^6\,L_\odot}{L}\right)^{3/4}
\end{equation}
\cite{GraefenerVink2013}, would be higher.
For the lower luminosity, we would need a mass-loss rate of $10^{-5.4}\,M_\odot\,\mathrm{yr}^{-1}$ to reproduce the spectrum. The inferred mass from the hydrodynamically-consistent model would go down to $\sim$$23\,M_\odot$, very similar to M33WR 206 and WR\,2. A slight underprediction of \ion{O}{vi}\,$3811\,$\AA\ from our model could be compensated by a very small increase in the oxygen abundance. Thus, while we cannot rule out dilution for this target and consequently the luminosity and the dependent parameters have to be considered more uncertain than for the other three studied stars, this is only a scaling effect. The general conclusions about the abundances, the evolutionary status, and the production of significant ionizing flux beyond the \ion{He}{ii} edge all remain untouched and can thus be considered robust.

\subsection*{Spectral analysis of the WN3 star WR\,46}

In addition to the studied WN2 and WN/WO stars, \ion{O}{vi}\,$\lambda\lambda$3811/3834 is also seen in the spectrum of the Galactic WN3 star WR\,46, where it appears as two peaked emission lines. With this emission as well as a prominent \ion{O}{V}\,$\lambda$1371 profile, the star shares some similarities to the WN2 and WN/WO stars, although the general weak-lined appearance with intrinsic self-absorption features in the emission lines in the Pickering series is actually more similar to weak-lined WN3h stars such as WR3 and those in the SMC. However, in contrast to those objects, WR\,46 is completely hydrogen free. Periodic behavior has been measured photometrically and spectroscopically \cite{Veen+2002a,Veen+2002b}, but despite early binarity claims \cite{Niemela+1995} a fully coherent pattern typically seen in WR+O binaries is missing. In particular, the radial velocities sometimes show a cyclic variability, but at other times do not vary at all, thus leading to the interpretation that this star is subject to non-radial pulsations \cite{Veen+2002c}. Short-term variability was also observed in the FUSE observations of the \ion{S}{vi}\,$\lambda\lambda$933/944 and \ion{O}{vi}\,$\lambda\lambda$1032/1038 wind lines \cite{Henault-Brunet+2011}. The X-ray emission shows variability on a timescale of about 7.9h \cite{Gosset+2011} and is high compared to the bulk of WR stars, but typical with respect to O-type stars \cite{Oskinova2005}. While stronger X-ray emission can be a signature of a colliding-wind binary, it is also known from weak-winded early-type WR stars, such as the WN3 stars WR\,3 and WR\,152 or the WO2 star WR\,142 \cite{Oskinova2005,Oskinova+2009,GuedelNaze2009}. Given that WR46 has a weaker wind and further been intensively monitored in the past without clear evidence for a luminous or dark companion, the wind-intrinsic X-ray scenario seems to be more likely here.

We perform a quantitative spectroscopic analysis with the PoWR code using available optical (ESO Faint Object Spectrograph at the ESO 3.6-metre telescope, PI Werner), high-resolution IUE (MA011, PI Willis), and FUSE (E113, PI St-Louis) spectra plus photometry from Gaia DR3 \cite{GaiaDR3}, 2MASS \cite{Cutri+2003}, and public WR narrow-band values \cite{Torres-DodgenMassey1988}. The distance of $d = 2.258\,$kpc is taken from the literature \cite{Crowther+2023} and based on the Gaia DR3 parallax. From our spectral analysis, we can conclude that the star is in an interesting regime that has been marked as ``forbidden'' by a combination of structural and wind constraints \cite{Grassitelli+2018,Sander+2023} as no sufficient stellar wind could be launched from the hot iron bump. However, recent 3D radiation-hydrodynamic simulations show that instead of an expansion of the star to cooler temperatures, the hot iron opacity bump can also create a regime of radiation-driven turbulence, which would allow for a weaker wind-launching without a significant expansion \cite{Moens+2022}. Given the large wind transparency due to the low mass-loss rate, the observed semi-cyclic variability could potentially be the result of such a turbulent regime in the wind-launching region.

The combined UV and optical spectroscopy can be sufficiently reproduced with one single atmosphere model, including the optical \ion{O}{vi} emission, indicating that no (luminous) companion is necessary to produce the observed spectrum. The mass-loss rate of WR\,46 is even lower than those of the WN2 and WN/WO stars. While this is mainly due to the lower luminosity of the WR star, the resulting transformed mass-loss rate is also a bit lower with $-5.14$. The terminal wind speed of about $3000\,$km\,s$^{-1}$ is on the order of other hot, weak-lined WN stars in the Milky Way such as our analyzed WN2-star WR\,2 or the weak-lined WN3 star WR\,3, which does not show optical \ion{O}{vi} emission. Confirming earlier indications \cite{Crowther+1995b}, WR\,46 shows an unusual oxygen abundance for a WN star as it is impossible to explain the optical and UV oxygen profiles with an oxygen mass fraction similar or smaller than the carbon mass fraction. Instead, the oxygen mass fraction is about three to five times those of carbon, which is not untypical for massive stars that did not lose their complete hydrogen shell. Still, oxygen is about an order of magnitude lower than the solar oxygen abundance, meaning that most of its initial abundance must have been converted into nitrogen. Given that such a pattern is not observed in any of the WN2 or WN/WO stars, not predicted by current evolution models for hydrogen-free stars, and the total oxygen abundance remains much lower than in the WN/WO objects, it is highly unclear whether this object is actually in a transition towards a WO stage. Alternatively, the stellar surface might have been polluted with material that has not undergone full CNO mixing, for example due to prior mass transfer from a former companion.

\begin{figure*}[hp]
    \centering
    \includegraphics[width=0.99\textwidth]{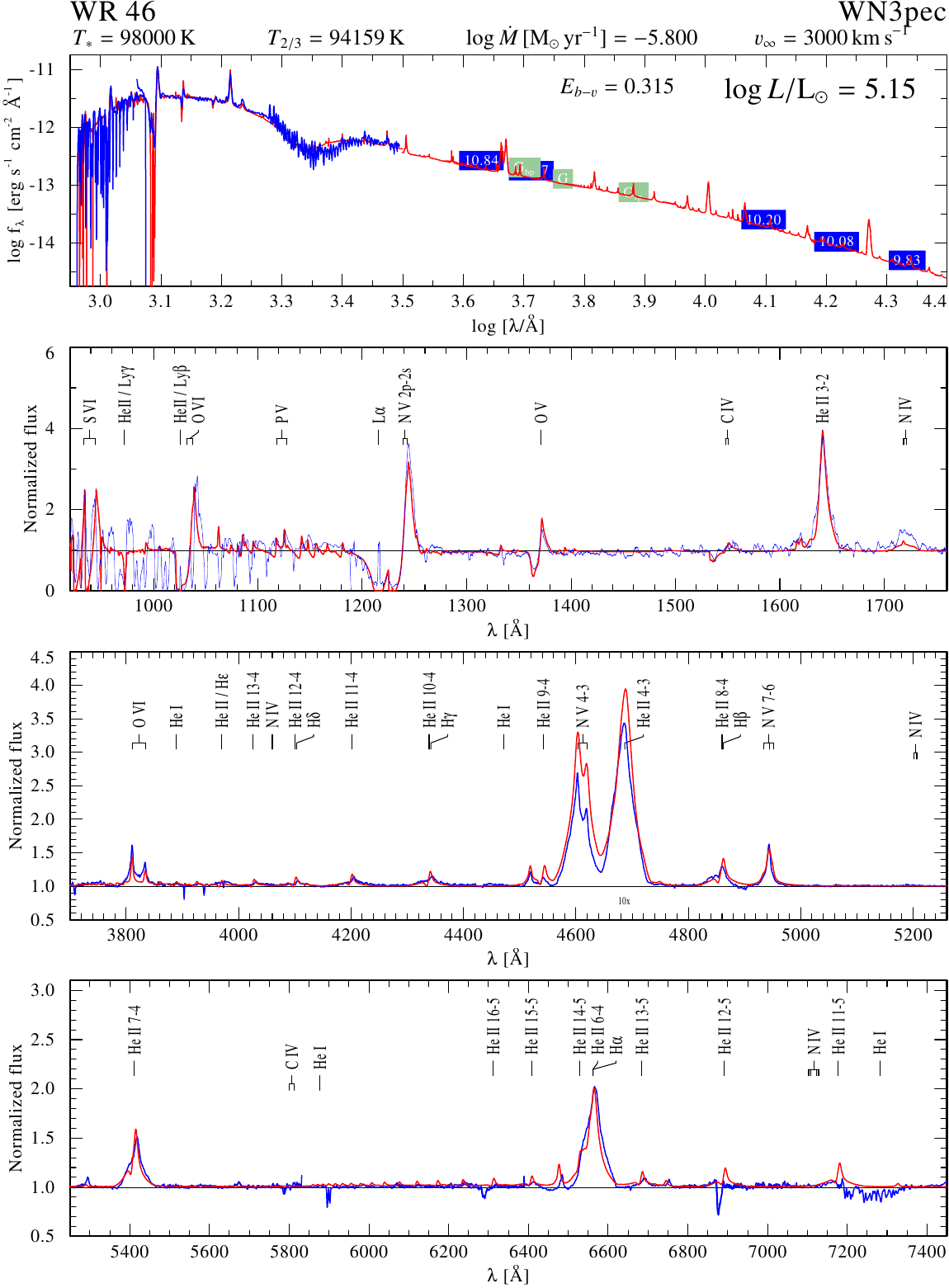}
    \caption{Observed data over-plotted with the best-fitting atmosphere model for WR\,46}
    \label{fig:WR46}
\end{figure*}

Given the uncommon oxygen abundance combined with an otherwise typical WN composition and the inaccessibility of a neon mass fraction due to the lower temperature, we do not consider WR\,46 in our main discussions of the WN to WO transition as it is not clear whether the current surface abundances already reveal a mixed layer right above the zone which underwent core-He burning. Moreover, the temperature of the object is notably cooler than those of the WN2 stars, meaning that even in the case of an increase of the carbon surface abundance, the current object might be classified as WC rather than a WO. Nonetheless, the star is a source of \ion{He}{ii} ionizing flux, albeit about 1.5 orders of magnitude lower than the WN2 and WN/WO stars due to the lower luminosity and cooler temperature. The full set of derived parameters, abundances, and ionizing fluxes is given in Table\,\ref{tab:wr46results} while the best-fit synthetic spectrum is compared to the available observational data in Fig.\,\ref{fig:WR46}.

\begin{table*}[ht!]
   \caption{Derived stellar parameters for WR\,46\label{tab:wr46results}. The assumed clumping parameters are $D_\infty = 10$ and $v_\mathrm{cl} = 100\,$km\,s$^{-1}$. }
   \centering
    \renewcommand{\arraystretch}{1.3}
    \begin{tabular}{l c p{1cm} l c}
      \hline 
         Parameter                   &               & &   Abundances   &  (mass fraction) \\ 
      \hline
       $T_{2/3}$ ($\tau=2/3$) [kK] &  $94^{+4}_{-4}$  &                     
                                                      &   $X_\mathrm{He}$      &  $0.984$     \\
       $R_{2/3}$ [$R_\odot$]       &   $1.41^{+0.20}_{-0.09}$   &   
                                                      &   $X_\mathrm{C}$     &   $1.0^{+1.0}_{-0.5} \cdot 10^{-4}$\\
       $\log L/L_\odot$            &   $5.15^{+0.05}_{-0.05}$  &
                                                      &   $X_\mathrm{N}$    &   $0.012^{+0.003}_{-0.002}$ \\
       $\log \left(\dot{M}\,[M_\odot\,\mathrm{yr}^{-1}]\right)$  &  $-5.85^{+0.05}_{-0.05}$  &
                                                      &   $X_\mathrm{O}$     &    $3.0^{+1.0}_{-1.0} \cdot 10^{-4}$ \\
       $\log \left(\dot{M}_\mathrm{t}\,[M_\odot\,\mathrm{yr}^{-1}]\right)$  & $-5.14^{+0.02}_{-0.02}$ &
                                                      &  $X_\mathrm{Ne}$    & $1.26\cdot 10^{-3}$   \\
       $v_\infty$ [km\,s$^{-1}$]      &   $3000^{+200}_{-400}$  & & $X_\mathrm{S}$    & $3.09\cdot 10^{-4}$   \\
       $L_\mathrm{mech}$\,[erg\,s$^{-1}$] &  $4.0 \cdot 10^{36}$  & & $X_\mathrm{Fe}$    & $1.4\cdot 10^{-3}$ \\
                                      &                & &                    &           \\
      \hline
         Ionizing photon rates     &               & &         &        \\ 
      \hline
       $\log \left(Q_\text{\ion{H}{i}}\,[\mathrm{s}^{-1}]\right)$    & $49.0^{+0.1}_{-0.1}$  &  &  $\log \left(Q_\text{\ion{Ne}{iii}}\,[\mathrm{s}^{-1}]\right)$  &  $45.8^{+0.3}_{-0.2}$  \\
       $\log \left(Q_\text{\ion{He}{i}}\,[\mathrm{s}^{-1}]\right)$   & $48.8^{+0.1}_{-0.1}$  &  &  $\log \left(Q_\text{\ion{Ne}{iv}}\,[\mathrm{s}^{-1}]\right)$   &  $41.8^{+0.3}_{-0.2}$  \\
       $\log \left(Q_\text{\ion{He}{ii}}\,[\mathrm{s}^{-1}]\right)$  & $46.7^{+0.2}_{-0.2}$  &  &  $\log \left(Q_\text{\ion{Ne}{v}}\,[\mathrm{s}^{-1}]\right)$    &  $38.4^{+0.4}_{-0.4}$  \\
      \hline       
    \end{tabular}
\end{table*}

\subsection*{Comparison of the observed HRD positions with evolutionary models}

\begin{figure*}[htbp]
    \centering
    \includegraphics[width=0.9\textwidth]{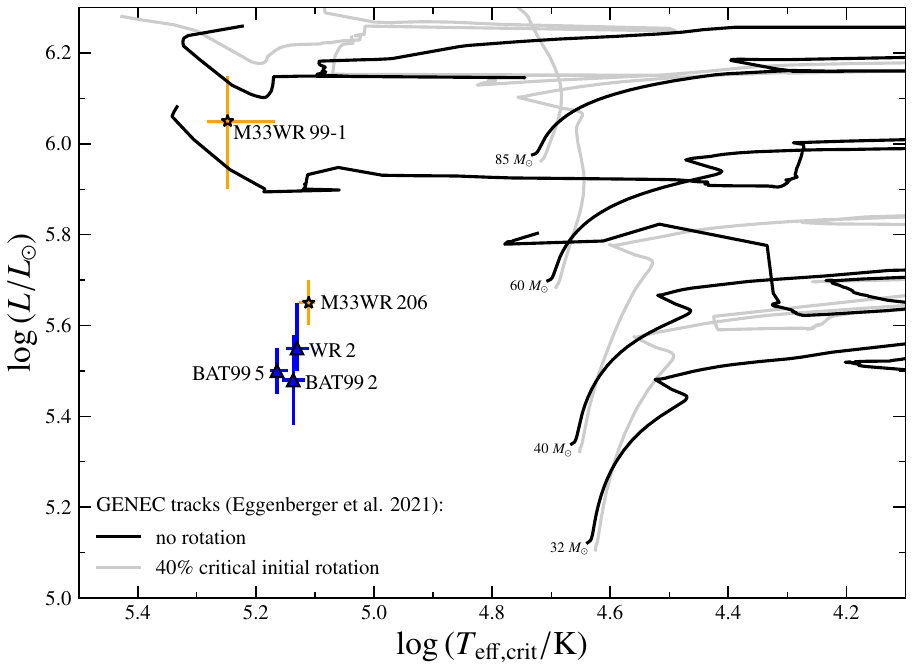}
    \caption{Hertzsprung Russell Diagram (HRD) with the positions of the newly discovered WN/WO stars (yellow) and the three other WN2 stars (blue) compared to the current generation of GENEC stellar evolution models with an LMC-like metallicity \cite{Eggenberger+2021}.} 
    \label{fig:hrd-genec2021}
\end{figure*}

\begin{figure*}[htbp]
    \includegraphics[width=0.9\textwidth]{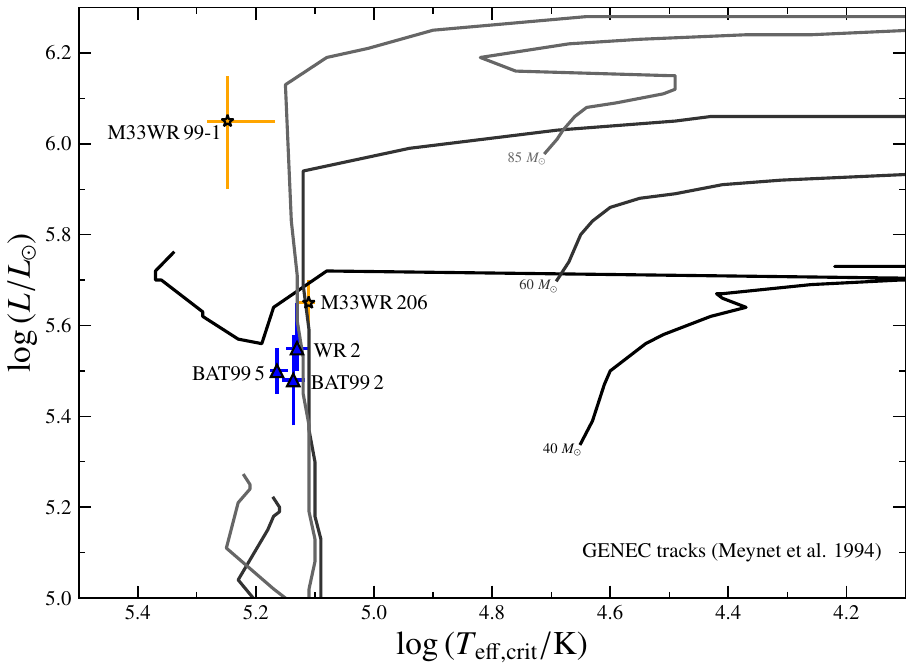}
    \caption{Same as EDF.\,\ref{fig:hrd-genec2021}, but showing the earlier tracks from \cite{Meynet+1994} with enhanced mass loss often employed in Starburst99 population synthesis models.} 
    \label{fig:hrd-genec1994}
\end{figure*}

In Fig.\,\ref{fig:hrd-genec2021}, we show the best-fitting HRD positions when comparing our WN/WO and WN2 stars with the current set of published GENEC stellar evolution models for LMC metallicity \cite{Eggenberger+2021}, which is closest in metallicity with respect to our determined abundances. For the models without rotation, only the HRD position for the more luminous M31WR\,99-1 is reasonably reached with the star being located between the tracks with initially $60$ and $85\,M_\odot$. None of the tracks reach the domain of any of the other WN2 and WN/WO stars. The $40\,M_\odot$ track does move away from the red supergiant regime and eventually reaches a hydro-free WN-like surface composition, but the effective temperatures only reach around $50\,$kK, much lower than observed for our studied targets. 
The comparison is even worse when considering the models with initially 40\% critical rotation (gray lines in Fig.\,\ref{fig:hrd-genec2021}) as no models below $85\,M_\odot$ even reach the desired temperature regime. GENEC models are commonly included in population synthesis models such as Starburst99, though not necessarily the most recent generation is used. Earlier models from 1994 \cite{Meynet+1994} are also available and are characterized by significantly larger mass-loss rates than the more modern models. While among other missing improvements, the overall enlarged mass-loss rates are not observationally supported by modern studies of resolved massive star populations, they are still a common choice in population sythesis \cite{Saldana-Lopez+2023,Garner+2025,Gunawardhana+2025} as these tracks manage to reach hot, H-free WR stages also for lower initial masses as subsolar metallicity. In Fig.\,\ref{fig:hrd-genec1994}, we show that these old models indeed do reach the observed regime, but the applicability of these models remains highly questionable.

\begin{figure*}[htbp]
    \centering
    \includegraphics[width=0.3\textwidth]{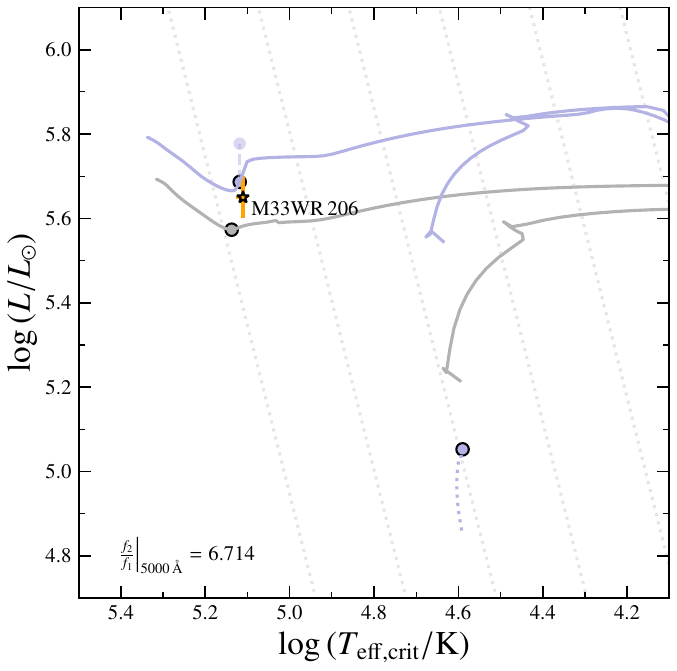}\mbox{\hspace{1em}}
    \includegraphics[width=0.3\textwidth]{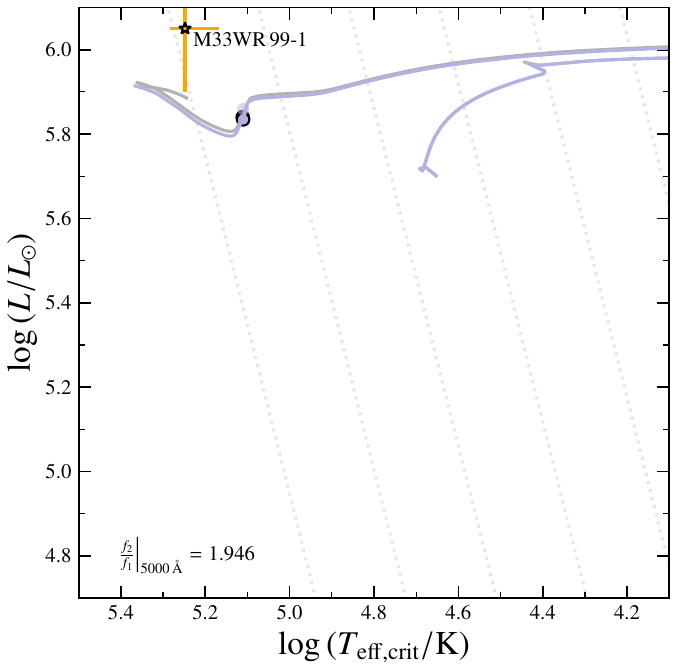}\\[1em]
    \includegraphics[width=0.3\textwidth]{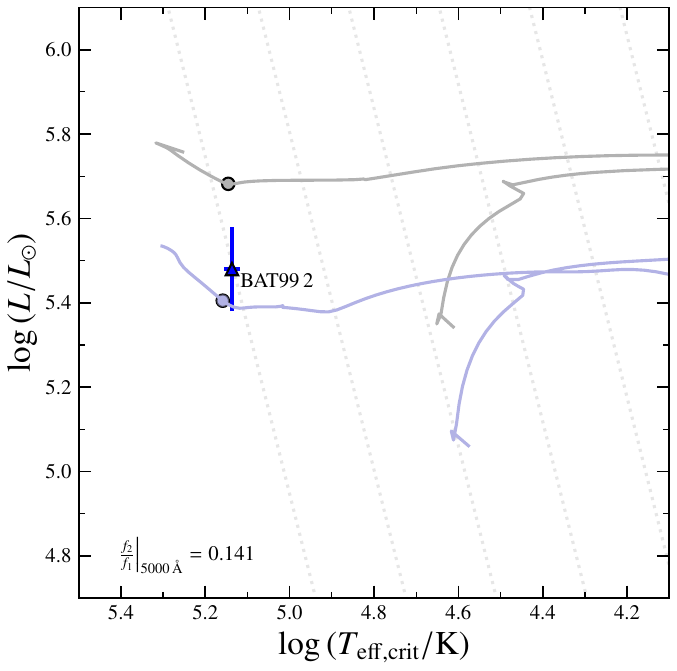}\hfill
    \includegraphics[width=0.3\textwidth]{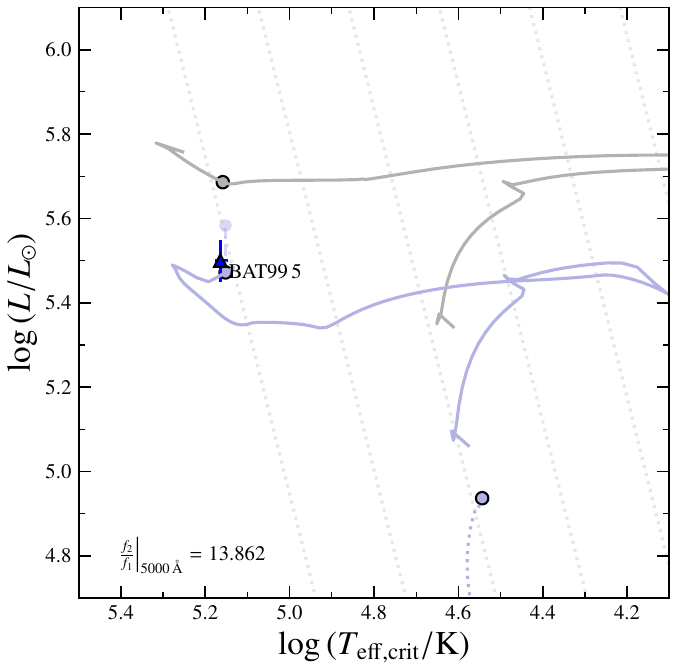}\hfill
    \includegraphics[width=0.3\textwidth]{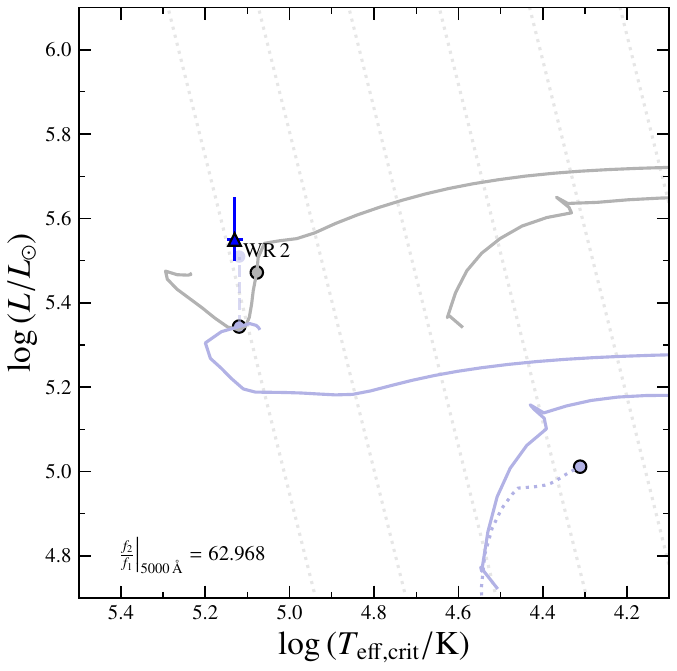}
    \caption{HRDs with the best-matching BPASS v2.2.1 \cite{Eldridge+2017,StanwayEldridge2018} single (grey) and binary (light purple) tracks for each of the analyzed WN/WO (yellow) and WN2 stars (blue). For binary solution, the secondary track is indicated by a dashed path. The combined luminosity of primary and secondary is indicated by a semi-transparent filled circle.} 
    \label{fig:hrd-bpass}
\end{figure*}

As most massive stars have companions and interact with them at least once during their lifetime, we also consider binary evolution models to explain the observed WN2 and WN/WO stars. Using a $\chi^2$-criterion involving $T_{2/3} = T_\mathrm{eff}(R_{2/3})$, $\log L$, and the surfance abundances of He, C, N, and O, we obtained the best-matching models from BPASS v2.2.1 looking for solutions where the primary has evolved into a WR star. In the five panels of Fig.\,\ref{fig:hrd-bpass}, the models with the minimum $\chi^2$ are shown, utilizing both single and binary evolution pathways. In general, the binary scenarios yield a better reproduction of the obtained HRD positions. However, we need to consider that in the selected BPASS models there is a secondary companion which should contribute to the total luminosity. As we do not see any indication for a luminous companion in our spectra, this would mean that the WR star needs to outshine the companion. To estimate this, we print the flux ratio at 5000\,$\AA$ using a simple blackbody assumption. While this is not very precise, all but one target (BAT99\,2) show numbers larger than unity, meaning that they clearly should be visible in the combined spectrum and usually provide the bulk of the optical flux, which is indeed common for known WR+O binary systems \cite{Shenar+2016}. Even the remaining BAT99\,2 with a $14\%$ contribution would usually be expected to be visible via absorption-line features. While this does by no means rule out a a binary evolution scenario, it implies that at least the applied sets of BPASS models are insufficient in explaining the WN2 and WN/WO stars. Alternative options considering binary evolution would be that our systems are either secondaries with a compact companion or mergers. Both of these scenarios cannot be ruled out, but would not be in line with current observations and interpretations. A compact companion, most likely a black hole (BH), would be limited to longer periods due to the absence of strong X-ray emission, while the stripping of the remaining non-compact star to its current stage likely requires a prior interaction with the compact object. Given the rarity of known WR+BH systems and the absence of dormant BHs for the SMC WN3 stars \cite{Schootemeijer+2024}, a merger option might look more preferential. However, the typically expected outcome of a merger is a blue supergiant, not a hot, H-free WR star. As argued for the SMC WN3 stars \cite{Schootemeijer+2024}, tailored scenarios with H-poor components are not impossible, but the more objects are known, the less likely becomes a channel requiring a high degree of fine tuning.

In principle, the observed HRD location of our analyzed stars could also be reached by quasi-chemically homogeneous evolution \cite{Maeder1987,Szecsi+2015,Kubatova2019,Szecsi+2025}. However, this would require rapid rotation, at least for the initial objects. With the currently weak winds in the observed WR stage, it would require sufficient fine tuning of the evolutionary path to spin down the stars prior to reaching the WR stage. This is not impossible, as recently shown in a demonstration setup trying to explain the single SMC WN3 stars \cite{Boco+2025}. Yet, there is no evidence for a current rapid spinning of the WN2 or WN/WO stars. The roundish emission lines of the WN2 star WR\,2 had led to such speculations \cite{Hamann+2006}, but a detailed investigation including spectropolarimetry could not find any indications for that \cite{Chene+2019}. Spectropolarimetric measurements of other lower-metallicity WR stars found evidence for high rotation only in a few targets, all of them having much stronger winds than our studied objects \cite{Vink+2017}. For the WN/WO stars, the transitional nature itself with the abundance mixture of He- and H-burning products rules out any recent chemically homogeneous evolution. While we cannot rule out that the studied stars might have initially evolved chemically homogeneous for a certain amount of time, the mixing of H- and He-burning products on the surface implies that chemical homogeneity most likely has been broken already before central H-burning was finished. We are also missing sufficient H-containing progenitors, which would be expected leftwards of the main sequence and should in principle be more numerous than the He-burning stars of our study. We thus do not explore chemically homogeneous evolution scenarios in our modeling efforts.

\begin{figure*}[htbp]
    \centering
    \includegraphics[width=0.99\textwidth]{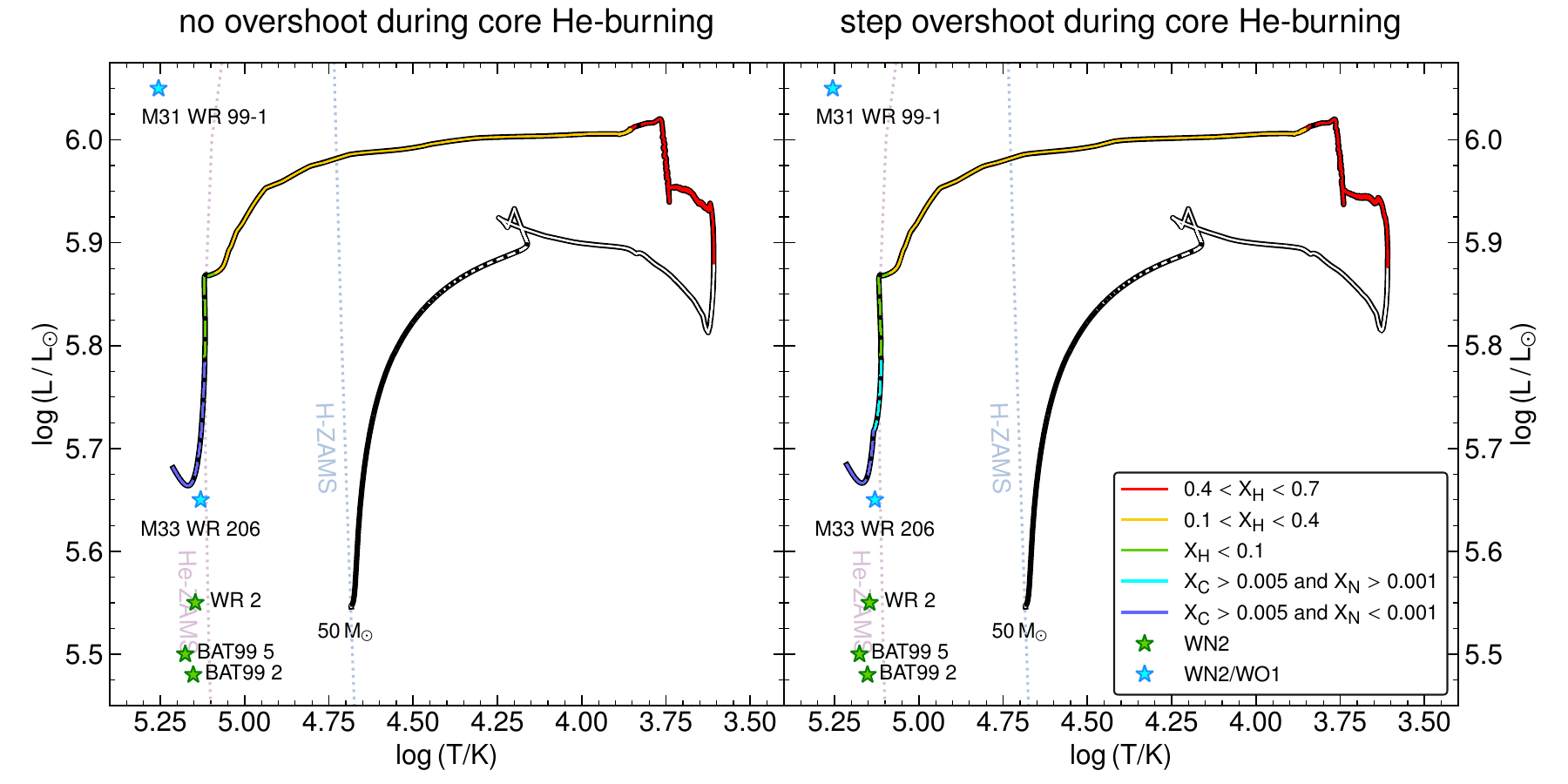}
    \caption{HRD showing the derived positions of our sample stars compared to tracks from custom MESA calculations without (left) and with (right) overshooting during the core-He burning. The models also use an increased mass loss in the cool star regime to remove a significant part of the envelope which otherwise employ the standard ``Dutch'' prescription. The surface composition is indicated by the track color and dots mark steps of 30\,kyr.} 
    \label{fig:hrd-mesa}
\end{figure*}

\begin{figure*}[htbp]
    \centering
    \includegraphics[width=0.99\textwidth]{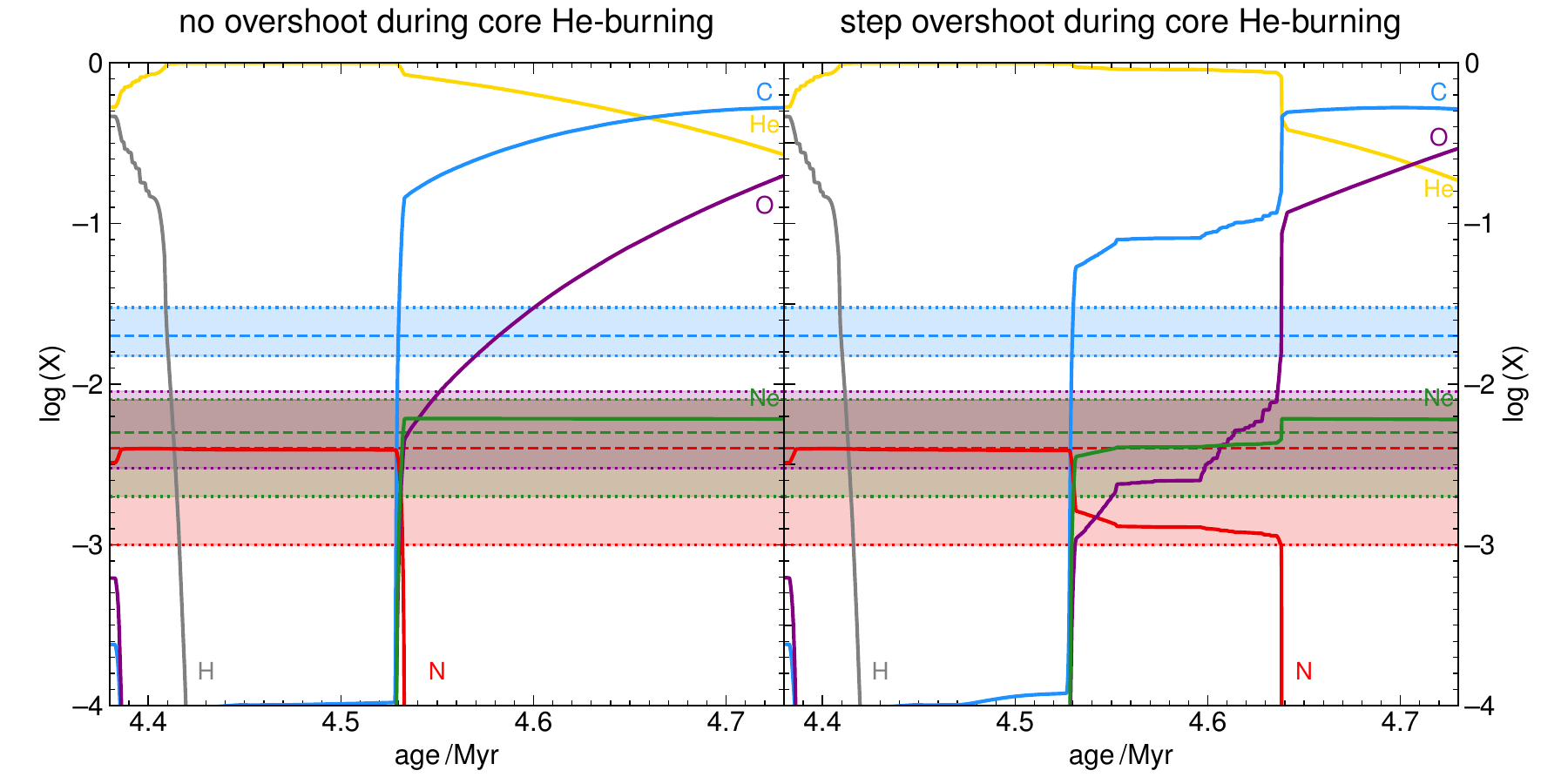}
    \caption{Surface abundances predictions as a function of age from MESA calculations without (left) and with (left) overshooting during the core-He burning. For comparison, the derived abundances for the WN/WO star M33WR\,206 are shown with the shaded regions indicating their uncertainties.} 
    \label{fig:wnwo-surface-abu}
\end{figure*}

\begin{figure*}[htbp]
    \centering
    \includegraphics[width=0.99\textwidth]{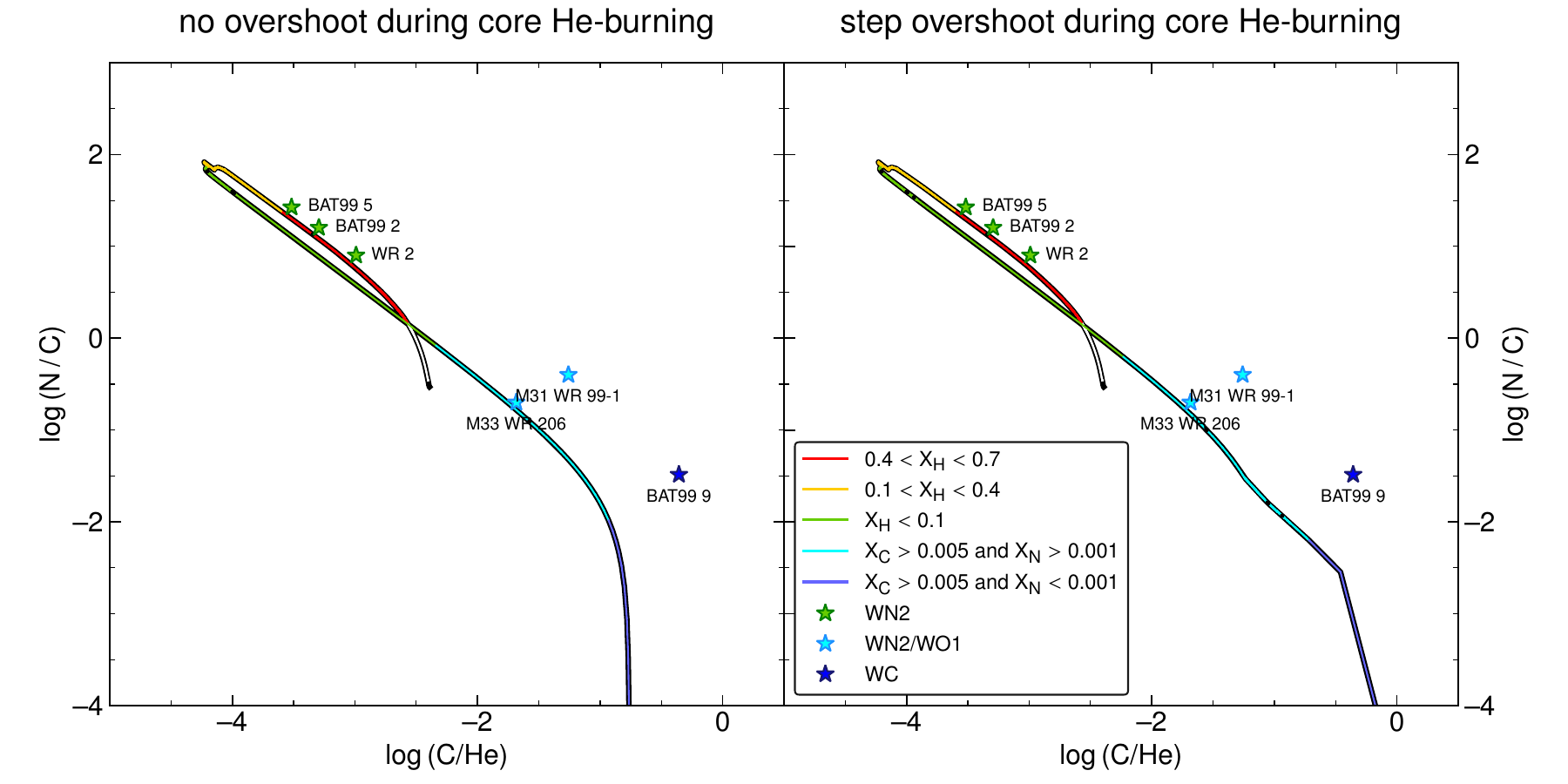}
    \caption{N/C versus C/He surface abundance ratios of our sample stars as well as the nitrogen-containing WC stars BAT99\,9 \cite{Hillier+2021} compared to the predictions from the MESA calculations without (left) and with (left) overshooting during the core-He burning. Similar to the HRD plot in Fig.\,\ref{fig:hrd-mesa}, the surface composition is indicated by the track color and dots mark steps of 30\,kyr.} 
    \label{fig:wnwo-surface-ratio}
\end{figure*}

Beside the HRD positions, the obtained surface abundances for the WN2 and WN/WO stars (cf.\ Table\,\ref{tab:results}) provide important properties which should be reproduced by the evolution models. Most models of massive star evolution, including the old and the new GENEC model grids \cite{Meynet+1994,Ekstroem+2012,Eggenberger+2021} or the BPASS models \cite{Eldridge+2017} predict a very rapid transition from the WN to the WC stage caused by the removal of the nitrogen-enriched envelope which then reveals layers enriched in carbon and oxygen produced during He burning. 
The occurrence of a layer that is both enriched in nitrogen and carbon/oxygen  implies that there is a slow-down or prevention of efficient convective mixing, possibly due to gradients in the mean molecular weight, which leads to a transition layer that is enriched in nitrogen due to the CNO cycle as well as material from the He-burning convective core \cite{Langer1991}.
We illustrate this in Fig.\,\ref{fig:hrd-mesa}, where we calculate fiducial MESA tracks at LMC metallicity. We employ a modified version of the ``Dutch'' \cite{deJager+1988,Vink+2001,NugisLamers2000} mass-loss scheme where we increase the mass loss in the cool regime to force the models to return back to the blue at LMC metallicity. While not aiming at a precise reproduction of the observed HRD positions, the tracks reach the general parameter regime of the observed WN2 and WN/WO stars. Yet, the surface abundances of the WN/WO stars can only be reached for a significant amount of time, if overshooting is also considered for the core-He burning core, which is usually not accounted for in published stellar evolution model grids. As further illustrated in Fig.\,\ref{fig:wnwo-surface-abu}, the overshooting prevents the otherwise almost immediate increase of the carbon surface mass fraction from $10^{-4}$ to more than $10\%$ and also delays the increase of the oxygen abundance. The duration of the WN/WO transition stage in Fig.\,\ref{fig:hrd-mesa} depends both on the prior evolution, the assumed mass-loss rate during the WR stage, and the overshooting assumptions. While reproducing the precise abundances of the observed stars is beyond the scope of this work, we show the general behaviour in the C/N versus C/He plane in Fig.\,\ref{fig:wnwo-surface-ratio}. For comparison, we also include the recently analyzed nitrogen-containing WC star BAT99\,9 \cite{Hillier+2021}. While there seems to be a shift with respect to the overall nitrogen level, the general trend is nicely reproduced. The time-indicating black dots again illustrate that the additional overshooting is necessary to increase the duration of the transition stage where both nitrogen as well as carbon and oxygen are seen at the surface. Our calculations thus show that future massive-star evolution models should no longer neglect the inclusion of overshooting in the core-He burning stage. Our WN/WO stars as well as known WN/WC will be important benchmarks and set time constraints on the duration of this transition stages, which by early WN/WC detections was estimated to be around $10\,000$ years \cite{Crowther+1995}, but could easily be an order of magnitude higher for particular evolution paths as indicated by our MESA calculation.

Despite the similarities of the WN/WO stars to WN/WC stars and the (partial) success of some evolution models to reproduce the HRD positions and abundances, the WN/WO and WN2 stars are fundamentally different than the WRs predicted by the evolution models. In published massive-star evolution modelling efforts \cite{Ekstroem+2012,Grasha+2021,Byrne+2025}, those tracks which reach the domain covered by early-type WR stars are assuming high current wind mass-loss rates for the predicted objects, just as observed for example for our comparison objects WR\,6 and BAT99\,52 in Fig.\,3. When the old GENEC models and our MESA calculations reach the observed HRD positions of M33WR\,206, WR\,2, BAT99\,2, and BAT99\,5, they too assume a strong mass-loss rate on the order of about $10^{-4}\,M_\odot\,\mathrm{yr}^{-1}$. This is about a factor of $40$ higher than observed. The mass-loss rates of the BPASS (primary) models are a bit lower, but with around $10^{-4.6}$ still an order of magnitude too high. This has dramatic consequences for both the feedback and the subsequent evolution. In contrast to the track predictions, the low current mass-loss rates imply that the WN2 and WN/WO stars will most likely not lose much more mass during their subsequent evolution. An exception is M33WR\,99-1, where the newer GENEC tracks \cite{Eggenberger+2021} yield a realistic mass-loss rate, but instead significantly overpredict the carbon and oxygen surface abundances.

New models that correctly reproduce the observed WN2 and WN/WO stars thus need a revision of the handling of WR-type mass loss going beyond the still common treatment which simply depends on the current luminosity and largely ignores the mass and radius-dependencies. Moreover, the strong WR mass-loss assumed in many evolution models is usually also an important factor in stripping the star to reach the early-type WR status in the first place, even if binary interactions also contributed significantly to a prior stripping of the envelope \cite{Pauli+2022}. This is also the case in our MESA demonstration models (cf.\ Fig.\,\ref{fig:hrd-mesa}), where the assumed strong WR mass loss helps to remove the remaining hydrogen envelope and afterwards leads to the formation of the downward evolution along the He main sequence. In particular, the latter is completely ruled out for our sample stars. Still, these stars must have lost a significant part of their envelope, including all of their hydrogen. While the current observational insights are insufficient to draw a clear conclusion on whether this must have been intrinsic or could involve multiplicity, it is clear that the mass loss cannot have happened at their current HRD position. The most likely scenario is thus a significant loss of mass at cooler temperatures, possibly hinting at a similar scenario than proposed for the single WN3ha stars in the SMC \cite{Schootemeijer+2024}.
 
\begin{figure*}[htbp]
    \centering
    \includegraphics[width=0.99\textwidth]{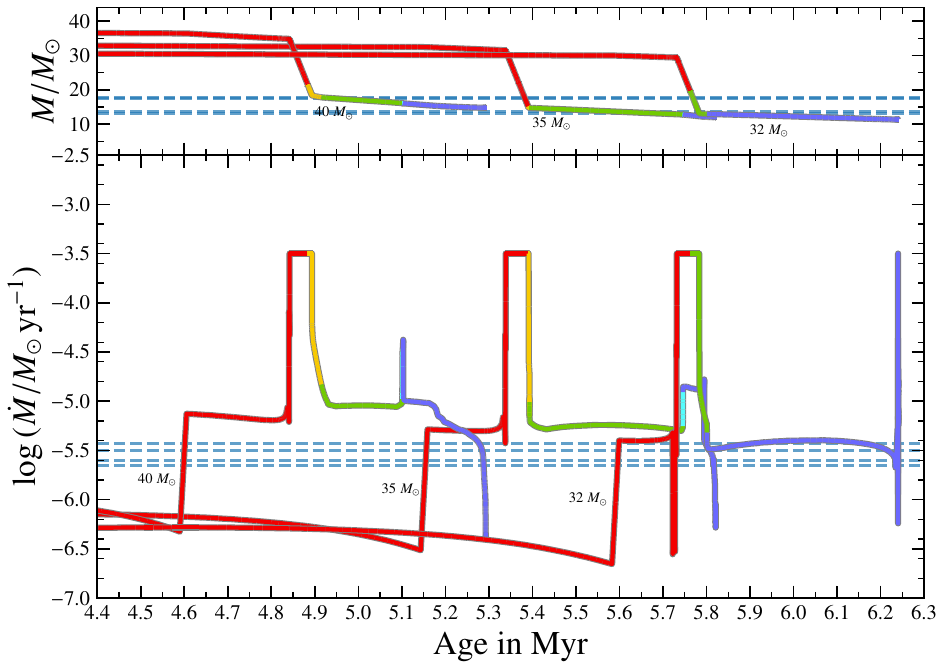}
    \caption{Mass (small upper panel) and mass-loss rate (lower panel) as a function of age for the custom GENEC models calculated with our adapted wind mass-loss scheme, focusing on the time beyond central hydrogen burning. Similar to the HRD plot in Fig.\,4, the surface composition is indicated by the track color and dots mark steps of 50\,kyr. For comparison, the derived masses and mass-loss rates of the three WN2 stars and M33WR\,206 are indicated as dashed light blue horizontal lines.} 
    \label{fig:genec-custom-mdot}
\end{figure*}

\begin{figure*}[htbp]
    \centering
    \includegraphics[width=0.99\textwidth]{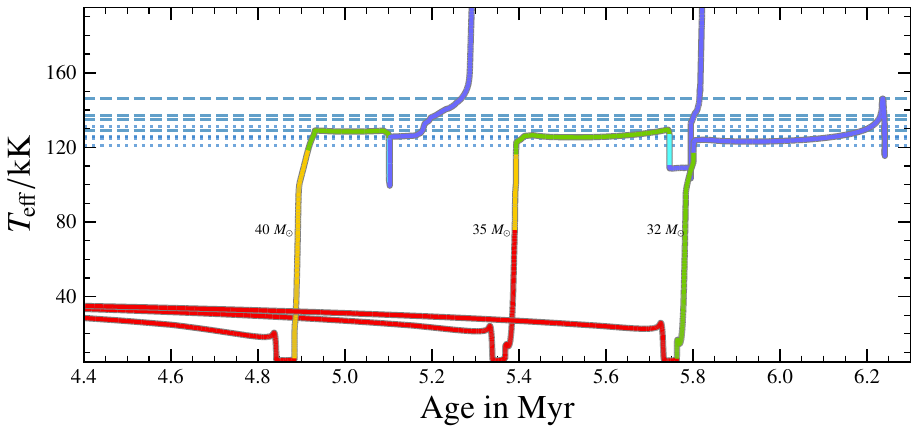}
    \caption{Effective temperature (without wind correction) as a function of age for the custom GENEC models calculated with our adapted wind mass-loss scheme, focusing on the time beyond central hydrogen burning. Similar to the HRD plot in Fig.\,4, the surface composition is indicated by the track color and dots mark steps of 50\,kyr. For comparison, the effective temperatures of the three WN2 stars and M33WR\,206 at the wind onset (dashed light blue) and the photosphere (dotted lighter blue) are indicated as horizontal lines.} 
    \label{fig:genec-custom-teff}
\end{figure*}

As a first demonstration, we calculate GENEC models that apply a large mass-loss rate of $\dot{M} = 10^{-3.5}\,M_\odot\,\mathrm{yr}^{-1}$ in the regime below 10\,kK. From both observational as well as theoretical considerations, a generally higher mass-loss rate on the main sequence is unlikely. When the models return to the blue, we employ the recent treatment from \cite{Sander+2023} for the WR stage for temperatures above 95\,kK without further adjustments. The resulting mass-loss rate and effective temperature in the hot WR stage align well with the observed values shown in Figs.\,\ref{fig:genec-custom-mdot} and \ref{fig:genec-custom-teff}. While the custom abundances. While the high mass loss in the cool stage is implemented as a continuous mass loss in our model, this does not necessarily imply a strong wind during this phase, but in principle could also arise from companion interaction(s) or eruptions. In this model, the star manages to lose most of its hydrogen envelope, forcing the evolution to return back to the hot side of the HRD, where the remaining hydrogen shell is lost. The whole process of losing the envelope happens early in core-He burning and takes less than 100\,kyr, while the rest of the core-He burning lifetime of at least 300\,kyr is spent in the hot, weak-winded WR stage as illustrated in Fig.\,\ref{fig:genec-custom-mdot}. The resulting set of GENEC tracks is compared to the obtained HRD positions -- using the GENEC temperatures not corrected for wind effects and the model-obtained effective temperatures at the wind onset -- in Fig.\,4 of the main manuscript. Time steps of 50\,kyr are indicated by dots along the tracks.

\begin{figure*}[htbp]
    \centering
    \includegraphics[width=0.99\textwidth]{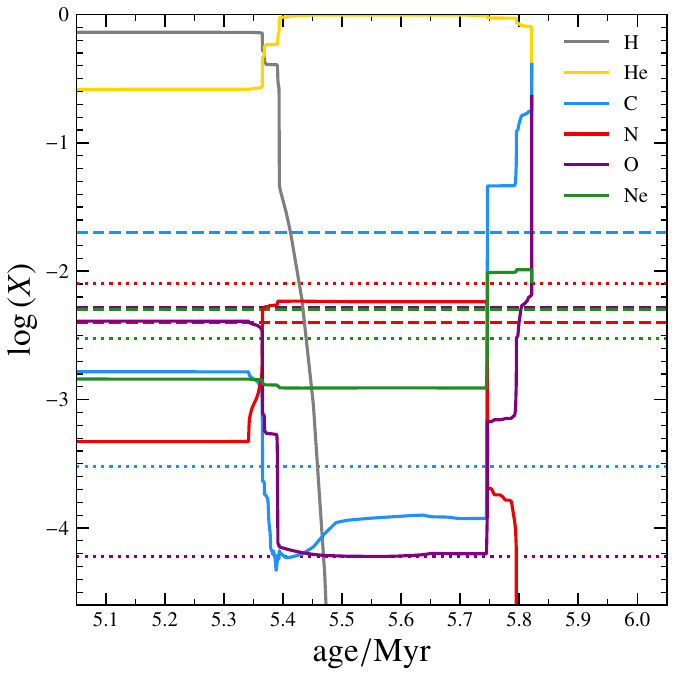}
    \caption{Surface abundances predictions as a function of age from the curstom GENEC model with an initial mass of $35\,M_\odot$ during the post-main sequence stages. For comparison, the derived abundances for the WN2 star BAT99-5 and the WN/WO star M33WR\,206 are shown as dotted and dashed horizontal lines, respectively.} 
    \label{fig:genec-custom-abu-m08}
\end{figure*}

A final issue to investigate in our custom GENEC models are the resulting surface abundances, which are exemplarily shown in Fig.\,\ref{fig:genec-custom-abu-m08} and have their main imprints also marked as coloring of the HRD tracks in Fig.\,4. With our adjusted mass-loss scheme, we generally obtain a WN-like surface composition after stripping the hydrogen shell. While we do not tweak individual elemental abundances and thus overpredict elements such as nitrogen or neon a bit, there is a qualitative agreement of this stage with the observed WN2 compositions. However, getting a noticeable timescale of the WN/WO stage is not as straight-forward as in the MESA models due to the different handling and parametrization of the mixing. As there is, for example, no separate overshooting parameter for the Helium core in GENEC, unlike in MESA, a general increase of the overshooting would also affect other mixing processes, which would spoil other surface abundances. While this can be seen as a deficiency in our models, this also underlines the potential of using WN/WO stars especially and envelope-stripped WR stars in general as benchmark targets for different mixing schemes in evolution codes. However, this requires a separate systematic study, which is beyond the scope of the current work.

\subsection*{Ionizing feedback estimates and exemplary comparisons}

The existence of the WN/WO stars provide evidence for an evolutionary path for WR stars at low metallicity where a large fraction of the central He burning is spent in a hot stage with comparably weak winds. While typical WR stars at higher metallicity have strong winds and do not yield \ion{He}{ii} ionizing flux, this path is characterized by the opposite behavior, reducing the resulting kinetic feedback, while providing a huge source of hard ionizing feedback. Taking the value of $Q_{\mathrm{He}\,\textsc{ii}} = 10^{48.4}\,\mathrm{s}^{-1}$, which represents three of our studied targets with rather moderate current masses between $15$ and $20\,M_\odot$, the production of \ion{He}{ii} ionizing photons is more than 10 times higher than the contribution by the high-mass X-ray binary M33\,X-7 \cite{Ramachandran+2022}. SSN\,9, the earliest O-star in the NGC\,346 cluster of the metal-poor SMC, has $Q_{\mathrm{He}\,\textsc{ii}} = 10^{46.5}\,\mathrm{s}^{-1}$ \cite{Rickard+2022}. While in such a cluster this O2 star is sufficient to dominate the \ion{He}{ii} ionizing flux, this is almost two orders of magnitude lower than our WN2 and WN/WO stars, despite these being about 0.6\,dex\ lower in total luminosity than SSN\,9. This means that even when accounting for the \ion{He}{ii} ionizing photon budget over the total lifetime of the different evolution stages, which for our WR stage is only about one order of magnitude shorter than that of the O2 star, WR stars following the outlined evolutionary WN2$\rightarrow$WN/WO path will still provide considerably more hard ionizing photons than a highly luminous (and thus also relatively rare) O2 star.
Notably, the difference in luminosity does affect the hydrogen ionizing photon flux, where the WN2 and WN/WO stars with $\log L/L_\odot \approx 5.55$ are yielding $Q_{\mathrm{H}\,\textsc{i}} = 10^{49.3}\,\mathrm{s}^{-1}$, which is a factor of four lower than the $10^{49.9}\,\mathrm{s}^{-1}$ from the O2 star in NGC\,346.

The skipping of the explicit WC stage in the case of a WN to WO transition also provides an interesting path to produce a WR population that would not yield a significant ``red`` (or ``yellow'') bump in an integrated galaxy spectrum. Such galaxies which only contain a ``blue'' bump as well as nebular \ion{He}{ii} are known both a low redshift (e.g., Mrk 1315 or SBS 0948+53  \cite{ShiraziBrinchmann2012}) and more recently with MACS J1149-WR1 observed by JWST also at higher redshift \cite{Morishita+2025}. Assuming a similar luminosity than derived for the bulk of our targets, commonly used early-type WC models in the literature provide $Q_{\mathrm{He}\,\textsc{ii}} = 10^{47.5}\,\mathrm{s}^{-1}$ for a very low metallicity of $0.02\,Z_\odot$ \cite{CrowtherHadfield2006}. Our objects provide an order of magnitude more ionizing photons per second and are found in a metallicity regime that is about an order of magnitude higher, which would also be more in line with the measured gas-phase oxygen abundances of the aforementioned galaxies ($\sim$$0.2\dots0.4\,O_\odot$).

Notably, the example of M33WR\,206 shows that even when the optical \ion{C}{iv} emission is small, the UV line can still be quite significant, which might pose an interesting future testcase for galaxies such as I Zw 18 or SBS 0335-052 that lack a coherent explanation of the strong nebular \ion{He}{ii} emission \cite{Kehrig+2015,Kehrig+2018,Mingozzi+2025}. With our standard luminosity of $\log L/L_\odot \approx 5.55$, about 52 stars of this kind would be required to explain the observed emission in I\,Zw 18 \cite{Kehrig+2015}, while SBS 0335-052 would require about 1200 such stars. These numbers would be reduced by a factor of four if we assume more luminous stars like M33WR\,99-1. While in particular the number for SBS 0335-052 is still high and thus such WR stars might not be the only contributor there \cite{Mingozzi+2025}, a number between 400 and 1000 WR stars would actually remove the previously claimed discrepancy between the necessary number of WR stars and the total mass of the six super star clusters found in SBS 0335-052 \cite{Kehrig+2018} with these cumulated objects even contributing $\sim$$10^{47.5}\,\mathrm{s}^{-1}$ of \ion{Ar}{V} ionizing photon flux.

\subsection*{Cross-check with nebular diagnostics and BPT location}

For BAT99\,2, earlier studies have been performed based on nebula spectra \cite{Naze+2003,Stock+2011}. The presence of the [\ion{Ar}{iv}] $4711.4/4740.2\,$\AA\ lines were already indicating a very high ionization status of the surrounding region. To ionize \ion{Ar}{iv}, photons with an energy above $59.81\,$eV ($\lambda < 207.3\,$\AA), which is higher than the \ion{He}{ii} ionization edge at $54.4\,$eV, are required. From the nebular measurements, \cite{Naze+2003} reported a \ion{He}{II} ionizing flux of $\log Q_\mathrm{He\,\textsc{ii}} = 47.6$. Our model yields a five times higher value ($48.3$) which is not in conflict as not all of the ionizing photons leaving the star need to be visible in the nebular emission. 

\begin{figure}[htbp]
    \includegraphics[width=\columnwidth]{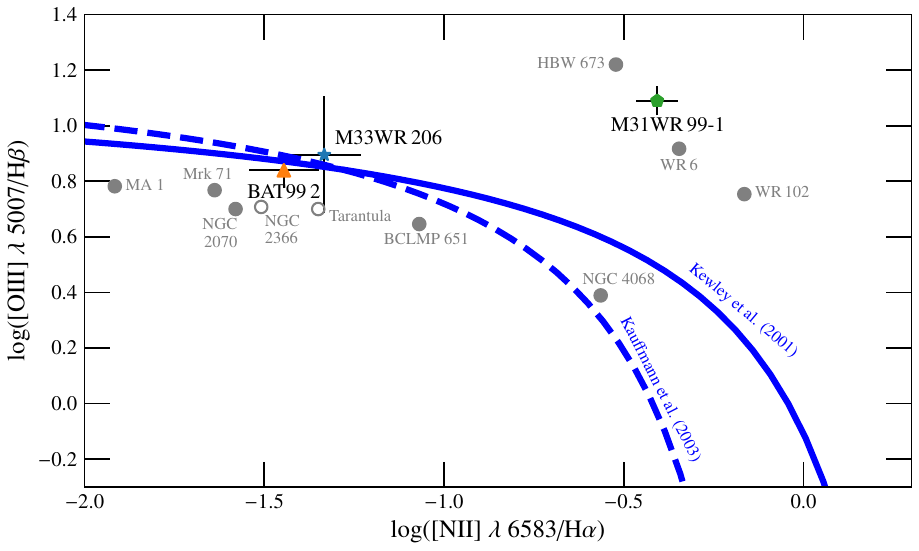}
    \caption{BPT diagram for the two WN/WO stars and the one WN2 star where nebular line diagnostics are available. Common diagnostic curves distinguishing between star-forming and AGN ionization regimes are plotted as dashed and solid blue lines. The gray symbols show the locations of other highly ionized regions for comparison. For both NGC2070 and Mrk 71, also the BPT locations of the larger area are plotted, i.e., the whole Tarantula region or respectively NGC 2366, as open circles.
    }
    \label{fig:bpt}
\end{figure}

Three of the five targets studied in this work are residing in \ion{H}{ii} regions for which nebular line information is available. From these, we can obtain the positions of the WN2 and WN/WO stars in the so-called BPT diagram \cite{Baldwin+1981}, which is shown in Fig.\,\ref{fig:bpt}. Common separation lines used to distinguish the dominant ionizing source of integrated populations are shown as dotted and dashed curves. The area below the curves in the lower left region is attributed to \ion{H}{ii} regions, meaning that the radiation of massive stars is considered to be the dominating ionizing source. The upper right area beyond the curves is instead attributed to ionization caused by an active galactic nucleus (AGN). The solid curve \cite{Kewley+2001} is also referred to as the ``maximum starburst line'', i.e., the reachable limit in terms of ionization from young stars, and based on then-current photo-ionization and population synthesis models. As this limit can already include galaxies also ionized by an AGN, the dashed curve \cite{Kauffmann+2003} marks a different separation with a less conservative estimate for the AGN influence, motivated also by empirical galaxy distributions. 

The dominant role in terms of ionizing feedback of our targets is further outlined by comparing the BPT locations to the well-studied star-forming regions, such as NGC2070 in the LMC Tarantula nebula \cite{CrowtherCastro2024} or the Mrk 71 region in the dwarf irregular galaxy NGC 2366 \cite{Micheva+2017}. These regions mark local analogues to so-called ``Green Pea'' (GP) galaxies characterized by extremely high ionization and occupy a similar location in the BPT diagram, as do our WN2 star WR 2 and the WN/WO star {M33WR\,206}. We further show the individual regions {MA 1}, {HBW 673}, and {BCLMP 651} in M33 \cite{Kehrig+2011} as well as object \#A in NGC 4068 \cite{Yarovova+2023}, where nebular He\,II emission has been spectroscopcially confirmed.

Within error bars, most of the plotted regions turn out to be in line with the star-forming regime. On the contrary, the WN/WO-star {M31WR\,99-1} is far beyond the traditional limit and located in the AGN regime. This underlines how highly ionizing sources are capable of exceeding the ionization limits commonly associated with a stellar origin. This behaviour is not unique to M31WR\,99-1, but also has been observed for ring nebulae associated with some WR stars \cite{Esteban+2016} such as the WO star WR\,102, the WN star WR\,6, or low-mass hot central stars of planetary nebulae \cite{Kniazev+2008}. Interestingly, {HBW 673} is also located in a similar regime, but marks one of the aforementioned examples of a \ion{He}{iii}-region that does not yet have a detected ionization source. Given the hard detectability of optically fainter WR stars such as {M31WR\,99-1} \cite{NeugentMassey2023}, so far undetected early-type WR stars may very well be the root of many so-far unexplained highly ionized regions such as {HBW 673}.

\begin{figure*}[hp]
    \centering
    \includegraphics[width=0.99\textwidth]{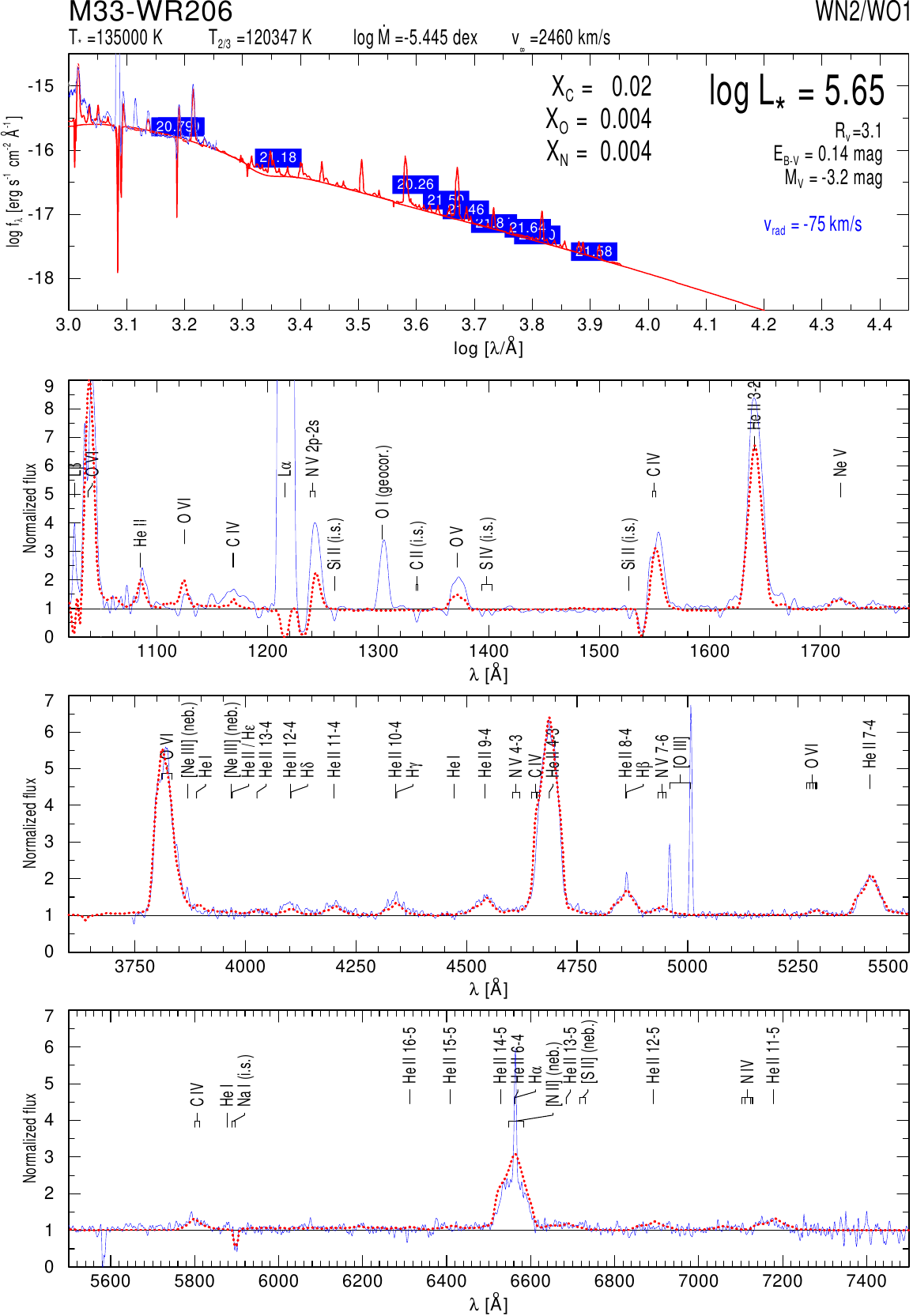}
    \caption{Observed data over-plotted with the best-fitting atmosphere model for M33WR\,206}
    \label{fig:M33WR206}
\end{figure*}

\begin{figure*}[hp]
    \centering
    \includegraphics[width=0.99\textwidth]{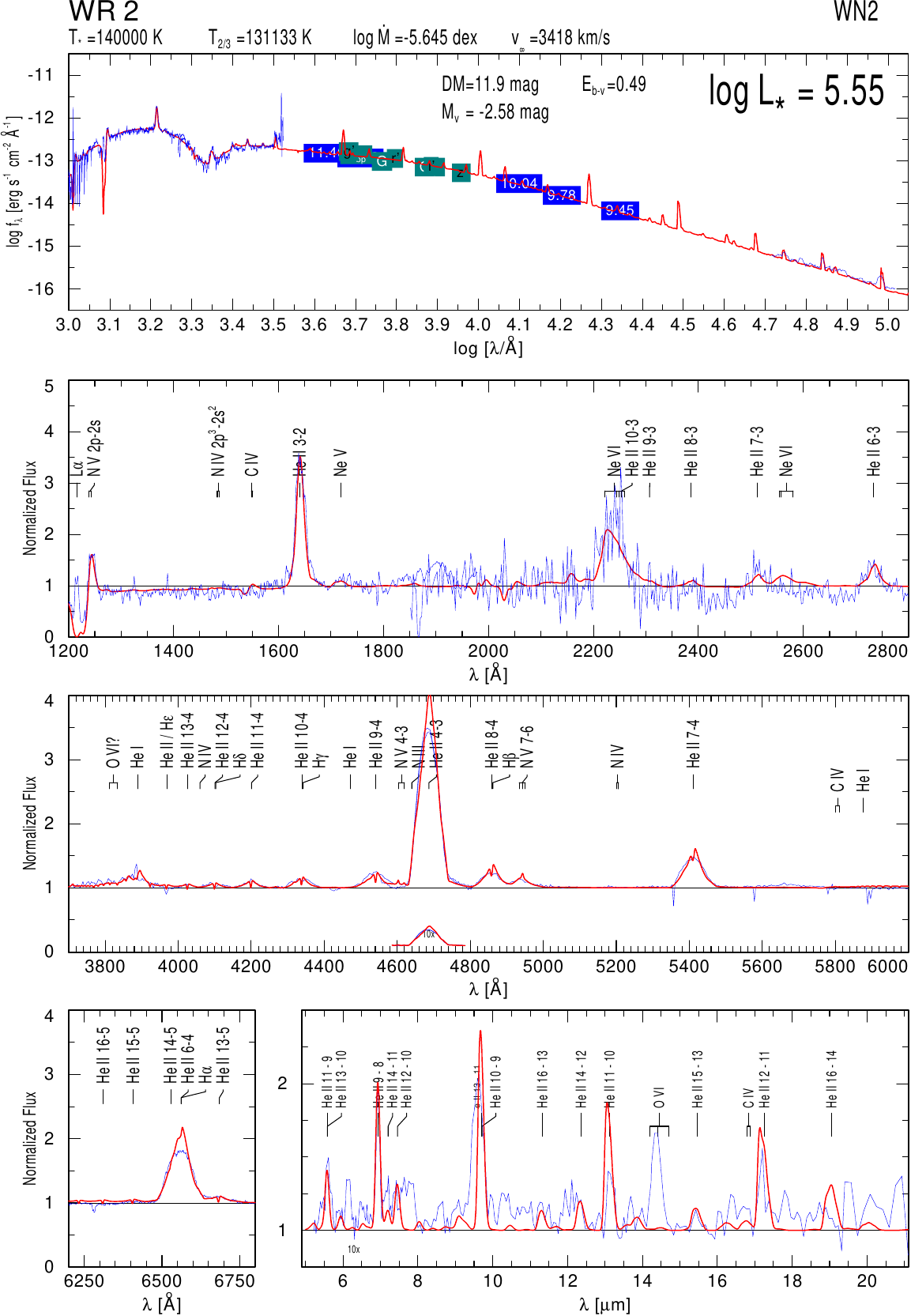}
    \caption{Observed data over-plotted with the best-fitting atmosphere model for WR2}
    \label{fig:WR2}
\end{figure*}

\begin{figure*}[hp]
    \centering
    \includegraphics[width=0.99\textwidth]{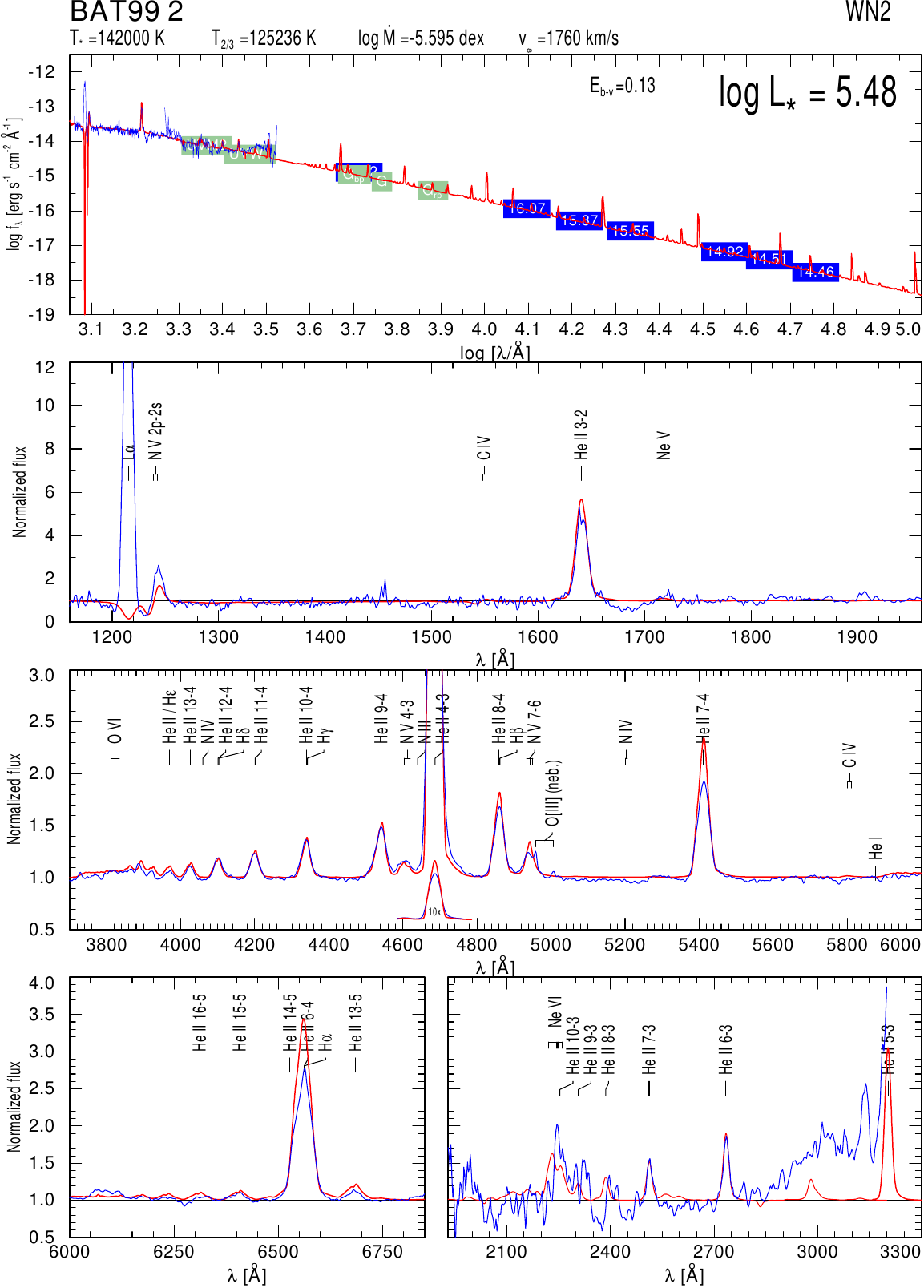}
    \caption{Observed data over-plotted with the best-fitting atmosphere model for BAT99\,2}
    \label{fig:BAT99-2}
\end{figure*}

\begin{figure*}[hp]
    \centering
    \includegraphics[width=0.99\textwidth]{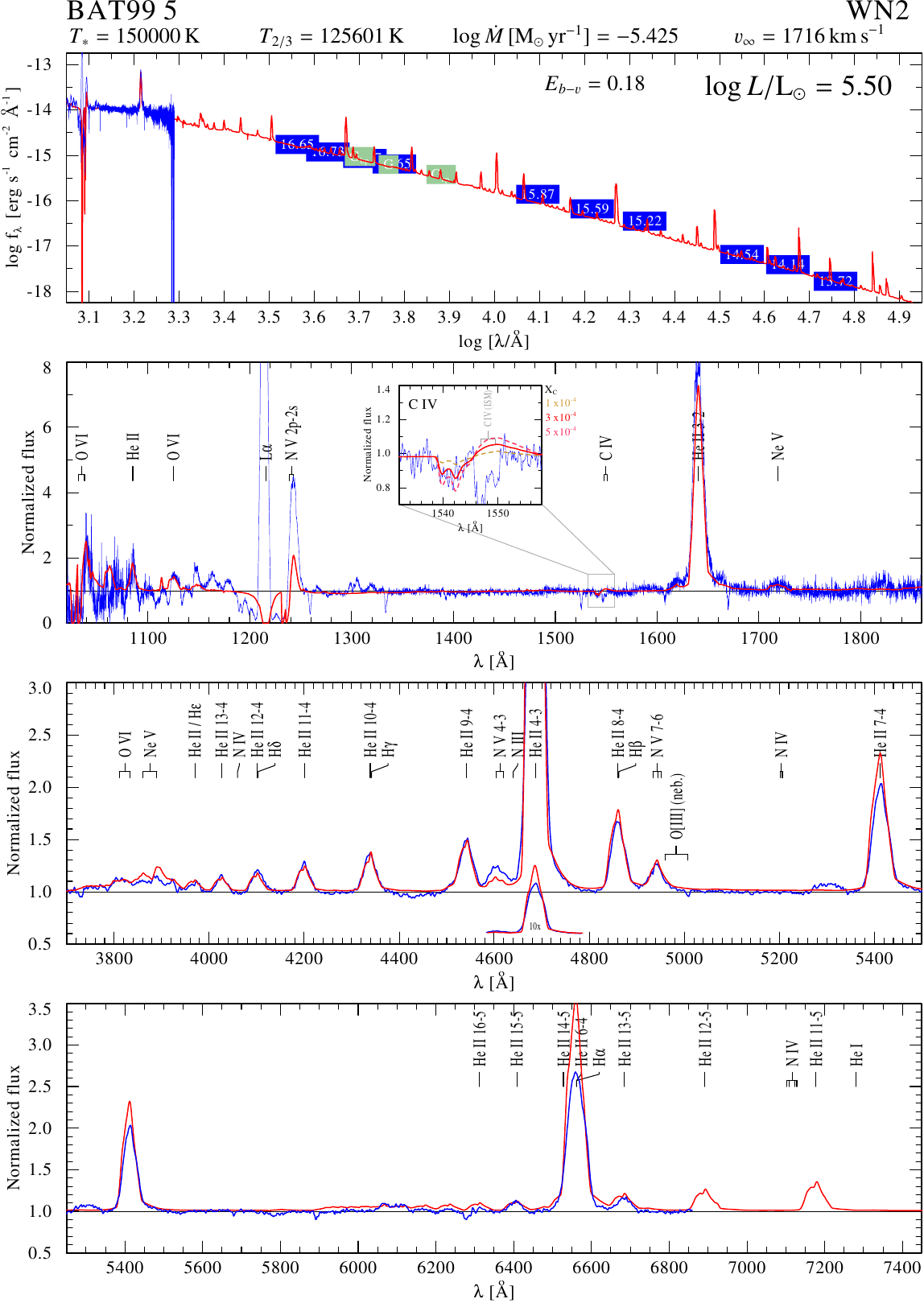}
    \caption{Observed data over-plotted with the best-fitting atmosphere model for BAT99\,5}
    \label{fig:BAT99-5}
\end{figure*}

\begin{figure*}[hp]
    \centering
    \includegraphics[width=0.99\textwidth]{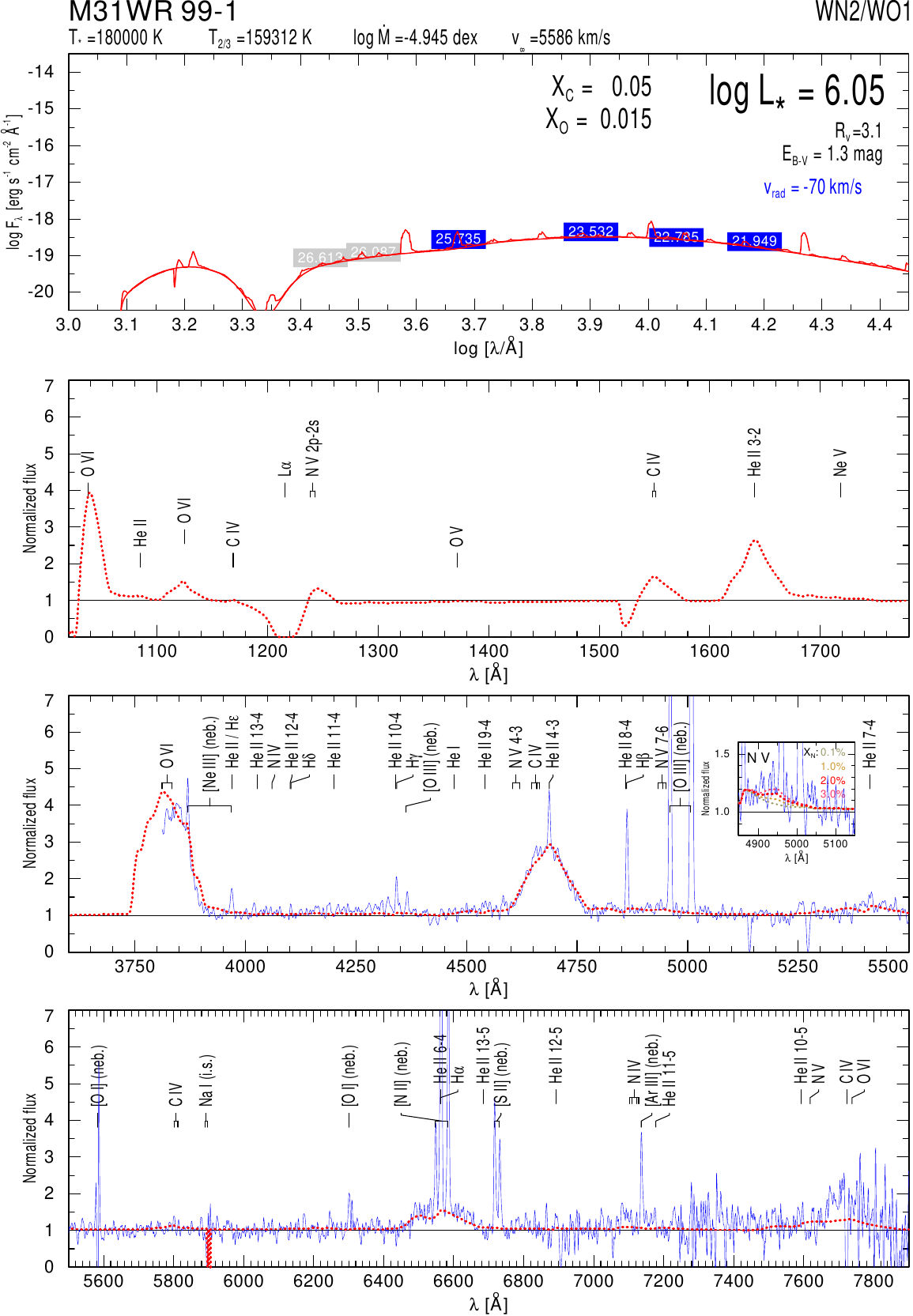}
    \caption{Observed data over-plotted with the best-fitting atmosphere model for M31WR\,99-1}
    \label{fig:M31WR99-1}
\end{figure*}

\begin{table}
   \caption{Number of levels and line transitions used in the PoWR$^\mathrm{\textsc{HD}}$
            models. The levels and lines listed for Fe -- a generic element that also
 			includes Sc, Ti, V, Cr, Mn, Co, and Ni -- correspond to superlevels and
			superlines \cite{Graefener+2002}\label{tab:datom}.}
   \centering
   \begin{tabular}{lcc c lcc c lcc}
      \hline 
      Ion    & Levels    & Lines   &  & 
      Ion    & Levels    & Lines   &  & 
      Ion    & Levels    & Lines  	  \\
    \hline
    \ion{He}{i}    &      35      &    595     & &   \ion{Mg}{iii}  &      10      &     45      & &    \ion{Ar}{vi}   &       9      &     36         \\ 
    \ion{He}{ii}   &      26      &    325     & &   \ion{Mg}{iv}   &      10      &     45      & &    \ion{Ar}{vii}  &      20      &    190         \\ 
    \ion{He}{iii}  &       1      &      0     & &   \ion{Mg}{v}    &      10      &     45      & &    \ion{Ar}{viii} &      11      &     55         \\ 
    \ion{C}{ii}    &       3      &      3     & &   \ion{Mg}{vi}   &      10      &     45      & &    \ion{Ar}{ix}   &      10      &     45         \\ 
    \ion{C}{iii}   &      40      &    780     & &   \ion{Mg}{vii}  &       1      &      0      & &    \ion{Ar}{x}    &       3      &      3         \\
    \ion{C}{iv}    &      25      &    300     & &   \ion{Al}{iii}  &      10      &     45      & &    \ion{K}{iii}   &      10      &     45         \\ 
    \ion{C}{v}     &       1      &      0     & &   \ion{Al}{iv}   &      10      &     45      & &    \ion{K}{iv}    &      23      &    253         \\ 
    \ion{N}{i}     &       3      &      3     & &   \ion{Al}{v}    &      10      &     45      & &    \ion{K}{v}     &      19      &    171         \\ 
    \ion{N}{ii}    &      38      &    703     & &   \ion{Al}{vi}   &      10      &     45      & &    \ion{K}{vi}    &      28      &    378         \\ 
    \ion{N}{iii}   &      30      &    435     & &   \ion{Al}{vii}  &       1      &      0      & &    \ion{K}{vii}   &       8      &     28         \\
    \ion{N}{iv}    &      38      &    703     & &   \ion{Si}{ii}   &      10      &     45      & &    \ion{K}{viii}  &       1      &      0         \\ 
    \ion{N}{v}     &      20      &    190     & &   \ion{Si}{iii}  &      10      &     45      & &    \ion{Ca}{iii}  &      10      &     45         \\ 
    \ion{N}{vi}    &       1      &      0     & &   \ion{Si}{iv}   &      10      &     45      & &    \ion{Ca}{iv}   &      24      &    276         \\ 
    \ion{O}{ii}    &      20      &    190     & &   \ion{Si}{v}    &      10      &     45      & &    \ion{Ca}{v}    &      15      &    105         \\ 
    \ion{O}{iii}   &      33      &    528     & &   \ion{Si}{vi}   &      10      &     45      & &    \ion{Ca}{vi}   &      15      &    105         \\
    \ion{O}{iv}    &      29      &    406     & &   \ion{Si}{vii}  &       1      &      0      & &    \ion{Ca}{vii}  &      10      &     45         \\ 
    \ion{O}{v}     &      54      &   1431     & &   \ion{P}{iii}   &      10      &     45      & &    \ion{Ca}{viii} &       1      &      0         \\ 
    \ion{O}{vi}    &      35      &    595     & &   \ion{P}{iv}    &      10      &     45      & &    \ion{Fe}{iii}  &       1      &      0         \\ 
    \ion{O}{vii}   &       1      &      0     & &   \ion{P}{v}     &      10      &     45      & &    \ion{Fe}{iv}   &      18      &     77         \\ 
    \ion{Ne}{ii}   &      10      &     45     & &   \ion{P}{vi}    &       1      &      0      & &    \ion{Fe}{v}    &      22      &    107         \\
    \ion{Ne}{iii}  &      18      &    153     & &   \ion{S}{iii}   &      10      &     45      & &    \ion{Fe}{vi}   &      29      &    194         \\ 
    \ion{Ne}{iv}   &      35      &    595     & &   \ion{S}{iv}    &      25      &    300      & &    \ion{Fe}{vii}  &      19      &     87         \\ 
    \ion{Ne}{v}    &      54      &   1431     & &   \ion{S}{v}     &      20      &    190      & &    \ion{Fe}{viii} &      14      &     49         \\ 
    \ion{Ne}{vi}   &      49      &   1176     & &   \ion{S}{vi}    &      22      &    231      & &    \ion{Fe}{ix}   &      15      &     56         \\ 
    \ion{Ne}{vii}  &      36      &    630     & &   \ion{S}{vii}   &       1      &      0      & &    \ion{Fe}{x}    &      28      &    170         \\
    \ion{Ne}{viii} &      77      &   2926     & &   \ion{Cl}{iii}  &       1      &      0      & &    \ion{Fe}{xi}   &      26      &    161         \\ 
    \ion{Ne}{ix}   &      20      &    190     & &   \ion{Cl}{iv}   &      24      &    276      & &    \ion{Fe}{xii}  &      13      &     37         \\ 
    \ion{Ne}{x}    &      15      &    105     & &   \ion{Cl}{v}    &      18      &    153      & &    \ion{Fe}{xiii} &      15      &     50         \\ 
    \ion{Ne}{xi}   &       1      &      0     & &   \ion{Cl}{vi}   &      23      &    253      & &    \ion{Fe}{xiv}  &      14      &     49         \\ 
    \ion{Na}{iii}  &       1      &      0     & &   \ion{Cl}{vii}  &       1      &      0      & &    \ion{Fe}{xv}   &      10      &     25         \\
    \ion{Na}{iv}   &      10      &     45     & &   \ion{Ar}{ii}   &      10      &     45      & &    \ion{Fe}{xvi}  &       9      &     20         \\ 
    \ion{Na}{v}    &      10      &     45     & &   \ion{Ar}{iii}  &      30      &    435      & &    \ion{Fe}{xvii} &       1      &      0         \\ 
    \ion{Na}{vi}   &      10      &     45     & &   \ion{Ar}{iv}   &      13      &     78      & &                   &              &                \\ 
    \ion{Na}{vii}  &       1      &      0     & &   \ion{Ar}{v}    &      10      &     45      & &                   &              &                \\    
    \hline
  \end{tabular}
\end{table}

\clearpage
\bibliography{literatur}

\section*{End Notes}

\subsection*{Acknowledgments}
This research is based on observations made with the NASA/ESA Hubble Space Telescope obtained from the Space Telescope Science Institute, which is operated by the Association of Universities for Research in Astronomy, Inc., under NASA contract NAS 5–26555. These observations are associated with programs 15822 (PI A.A.C.\ Sander) and 17426 (PI A.A.C.\ Sander). This paper benefited from discussions at the International Space Science Institute (ISSI) in Bern through ISSI International Team project 512 (Multiwavelength View on Massive Stars in the Era of Multimessenger Astronomy, PI L.M.\ Oskinova). R.R.L. and J.J. are members of the International Max Planck Research School for Astronomy and Cosmic Physics at the University of Heidelberg (IMPRS-HD).
A.A.C.S., R.R.L., and V.R. acknowledge support by the Deutsche Forschungsgemeinschaft (DFG, German Research Foundation) in the form of an Emmy Noether Research Group -- Project-ID 445674056 (SA4064/1-1, PI Sander). 
A.A.C.S. and V.R. further acknowledge support from the Federal Ministry of Education and Research (BMBF) and the Baden-Württemberg Ministry of Science as part of the Excellence Strategy of the German Federal and State Governments. J.J. acknowledges funding from the Deutsche Forschungsgemeinschaft (DFG, German Research Foundation) Project-ID 496854903 (SA4064/2-1, PI Sander). E.R.H. and J.S.V. are supported by STFC funding under grant number ST/V000233/1. L.M.O. is thankful for the funding provided by the DFG grant 443790621. D.P. acknowledges financial support by the Deutsches Zentrum f\"ur Luft und Raumfahrt (DLR) grant FKZ 50OR2005. 
R.H. acknowledges support from the World Premier International Research Centre Initiative (WPI Initiative), MEXT, Japan, the European Union’s Horizon 2020 research and innovation programme (ChETEC-INFRA, Grant No. 101008324) and the IReNA AccelNet Network of Networks (National Science Foundation, Grant No. OISE-1927130). This article is based upon work from the ChETEC COST Action (CA16117). I.M. acknowledges support from the Australian Research Council (ARC) Centre of Excellence for Gravitational Wave Discovery (OzGrav), through project number CE230100016.

\subsection*{Author contributions} A.A.C.S. developed the hypothesis, developed the hydrodynamically-consistent atmosphere PoWR code branch used in this work, calculated the final atmosphere models, and wrote most of the manuscript.  
R.R.L. calculated hydrodynamically-consistent atmosphere models, performed the spectral re-classification, and contributed to the manuscript.
J.J. calculated related GENEC evolution models and contributed to the interpretation of the objects.
E.R.H. provided WR star structure models and contributed to the discussion about their evolutionary nature.
R.H. contributed to the manuscript, in particular with respect to the evolutionary interpretation and the discussion of the abundances.
L.M.O. contributed the X-ray investigation of the studied targets and commented on the manuscript.
D.P. calculated MESA evolution models, contributed to the discussion on the evolutionary status, and commented on the manuscript.
M.P. calculated early atmosphere models for one of the involved targets.
J.S.G. contributed to the manuscript and the discussion regarding the wider context of the targets.
W.-R.H. developed the core and large parts of the PoWR atmosphere code and contributed to the discussion.
I.M., V.R., and T.S. contributed to the discussion regarding the nature and impact of the studied objects. 
H.T. collected a considerable part of the underlying atomic data to the atmosphere models, contributed to the PoWR atmosphere code, and to the atmosphere analysis and discussion.
J.S.V. contributed to the discussion on the evolutionary status and the hydrodynamically consistent atmosphere models.

\subsection*{Competing interests:} The authors declare no competing interests.

\subsection*{Correspondence and requests for materials} should be addressed to A.A.C.\ Sander

\subsection*{Data availability statement}

All observational material employed in the current study is available publicly from the sources listed in the ``observational data'' subsection of the methods section. The resulting synthetic spectra from the atmosphere models which support the findings of this study are plotted in the extended data figures. The raw model data is available upon request from the corresponding author.

\subsection*{Code availability}

The standard branch of the PoWR atmosphere model code is described in \cite{Graefener+2002,Hamann+2003,Sander+2015} and available on \url{https://github.com/powr-code}. The customized version by the corresponding author is described in \cite{Sander+2017,Sander+2020,Sander+2023} and available on request. A brief conceptual overview is further given in the ``Stellar atmosphere modeling'' subsection of the methods section with the employed set of atomic data summarized in extended data Table\,\ref{tab:datom}.

\end{document}